%% file: main.tex
\documentclass[journal ]{new-aiaa}
\usepackage[utf8]{inputenc}
\usepackage{textcomp}

\usepackage{graphicx}
\usepackage{amsmath}
\usepackage[version=4]{mhchem}
\usepackage{siunitx}
\usepackage{longtable,tabularx}
\usepackage{setspace}
\usepackage{cleveref}
\usepackage{array}    
\usepackage{multirow} 
\usepackage{algorithm}
\usepackage[noend]{algpseudocode}
\usepackage{gensymb}
\usepackage{subcaption}
\usepackage{pdfpages}
\usepackage{pgfplots}
\usepackage{booktabs}      
\usepackage{siunitx} 

\title{Free-Placement Optimization of Ground Station Locations for Low-Earth Orbit Satellites}

\author{Grace Ra Kim\footnote{Corresponding Author, PhD Student, Department of Aeronautics and Astronautics, Stanford, CA, AIAA Student Member, \texttt{gkim65@stanford.edu}}, Duncan Eddy\footnote{Postdoctoral Researcher, Department of Aeronautics and Astronautics, Stanford, CA, \texttt{deddy@stanford.edu}}, Vedant Srinivas\footnote{Undergraduate Student, Department of Computer Science, Stanford University, Stanford, CA, \texttt{vedants8@stanford.edu}}, and Mykel J. Kochenderfer \footnote{Associate Professor, Department of Aeronautics and Astronautics, Stanford, CA, AIAA Associate Fellow, \texttt{mykel@stanford.edu}}}
\affil{Stanford University, Stanford, CA, 94305}

\begin{document}

\maketitle

\begin{abstract}

Rapidly expanding low Earth orbit satellite constellations are placing increasing demands on terrestrial ground networks, motivating the development of more efficient ground station network designs. Current approaches select sites from predefined locations, limiting optimization to existing infrastructure and constraining performance. In contrast, free-placement optimization operates over a continuous spatial domain on Earth, broadening the search space and allowing higher-throughput configurations at the cost of potentially requiring new infrastructure deployment. In this work, we introduce SCORE (Sequential Cyclic Optimization via Refinement \& Evaluation), a two-stage free-placement method for ground station design. SCORE combines sequential coordinate selection with cyclic refinement to manage high-dimensionality, non-convexity, and local minima that challenge global optimizers. We benchmark SCORE against one-shot methods such as differential evolution (DE) and integer programming approaches using locations from Kongsberg Satellite Services and the World Teleport Association. Tests across two commercial Earth observation constellations (Capella Space and ICEYE) and one synthetic Walker-Star constellation show that SCORE requires up to 5$\times$ fewer function evaluations to converge relative to DE while improving downlink throughput by up to 13\%. Compared to fixed-site methods, unconstrained SCORE achieves up to 15\% greater total downlink, establishing a strong empirical performance benchmark for flexible placement; infrastructure-constrained SCORE retains over 92\% of this gain while restricting placement to within proximity of existing fiber and power infrastructure. We also explore trade-offs between expanding existing stations and deploying new sites, informing future ground network design for operational constellations.

\end{abstract}


\input{00_Nomenclature}

\input{01_Introduction}
\input{02_test_probFormulation}

\input{03_Methods_FreeSelection}

\input{05_ExperimentalSetup}
\input{06_Results}
\input{010_Conclusion}
\input{011_Appendix}

\bibliography{references}

\end{document}

%% file: 00_Nomenclature.tex
\section*{Nomenclature}


{\renewcommand\arraystretch{1.0}
\noindent\begin{longtable*}{@{}l @{\quad=\quad} l@{}}

$n$ & number of selected stations \\
$\lambda$ & longitude [degrees] \\
$\phi$ & latitude [degrees]\\
$\mathcal{L}$ & set of ground station locations \\
$L$ & single ground station, coordinate $(\lambda,\phi)$\\
$\mathcal{S}$ & set of satellites in a constellation\\
$S$ & single spacecraft in constellation\\
$f$ & performance objective function\\
$f_{\text{data}}$ & data downlink maximization objective function\\

$T_{\text{opt}}$ & time period of mission duration [s]\\
$t_{\text{opt}}^{\text{start}}$ & start time of mission duration [s]\\
$t_{\text{opt}}^{\text{end}}$ & end time mission duration [s]\\
$T_{\text{sim}}$ & time period of simulation [s] \\
$t_{\text{sim}}^{\text{start}}$ & start time of simulation [s]\\
$t_{\text{sim}}^{\text{end}}$ & end time simulation [s]\\

$\mathcal{G}$ & set of all coordinate points on Earth \\
$\mathcal{T}$ & set of all coordinate points on Earth's terrestrial surface \\
$\mathcal{I}$ & set of feasible infrastructure on Earth's surface \\
$\mathcal{P}$ & set of predefined locations from existing Ground Station as a Service sites or teleport locations \\
$P$ & single ground station in $\mathcal{P}$ \\
$p$ & binary decision variable for selecting location $P$ \\
$\mathcal{C}_{\mathcal{L,S}}$ & contact windows between network $\mathcal{L}$ and constellation $\mathcal{S}$ \\
$\mathcal{C}_L$ & set of contacts associated with station $L$ \\
$\mathcal{C}_S$ & set of contacts associated with spacecraft $S$ \\
$\mathcal{C}_P$ & set of contacts associated with location $P$ \\
$C$ & single contact object \\
$c$ & binary decision variable for selecting contact $C$ \\
$L_{\text{dr}}$ & fixed ground station data rate [Gbps]\\
$S_{\text{dr}}$ & fixed satellite data rate [Gbps]\\
$C_{\text{dr}}$ & fixed contact data rate [Gbps]\\
$C_{\text{duration}}$ & duration of contact $C$ [s]\\

$C^{\text{start}}$ & start time of contact [s]\\
$C^{\text{end}}$ & end time contact [s]\\

$Q$ & penalty function\\
$Q_\mathcal{T}$ & penalty for land placement constraint \\
$Q_{\text{proximity}}$ & penalty for station proximity constraint \\
$Q_{\mathcal{I}}$ & penalty for infrastructure proximity constraint \\
$L_\mathcal{T}$ & single ground station site in $\mathcal{T}$ \\
$d_{\text{min}}$ & minimum distance between ground stations [km]\\
$d_{L_i,L_j}$ & distance between locations $L_i$ and $L_j$ [km]\\
$r_{\text{infra}}$ & maximum allowable distance from infrastructure [km] \\
$L_{\mathcal{I}}$ & closest point in $\mathcal{I}$ to station $L$ \\

$t_{\text{min}}$ & minimum contact duration for contacts [s]\\
$E_{\text{max}}$ & maximum number of cyclic evaluations in SCORE algorithm\\ 
$e$ & evaluation count in SCORE algorithm\\ 



$\mathbf{x}$ & one-dimensional decomposed coordinate list of $\mathcal{L}$ for Differential Evolution algorithm\\ 
$D$ & dimensionality of $\mathbf{x}$ for Differential Evolution algorithm\\ 
$d$ & random index between $1 \dots D$ for Differential Evolution algorithm\\
$NP$ & population size parameter for Differential Evolution algorithm\\ 
$F$ & mutation scale factor for Differential Evolution algorithm\\ 
$CR$ & crossover rate parameter for Differential Evolution algorithm\\ 
$\mathbf{u}$ & mutated vector of $\mathbf{x}$ for Differential Evolution algorithm\\ 
$G_{\text{max}}$ & maximum generations for Differential Evolution algorithm\\
$g$ & generation count in Differential Evolution algorithm\\
$\mathcal{X}$ & population of candidate solutions for Differential Evolution algorithm\\

$N_\rho$ & number of planes in Walker-Star constellation\\
$\rho$ & single plane in Walker-Star constellation\\
$\Omega$ & Right Ascension of Ascending Node [degrees]\\
$\Omega_\rho$ & Right Ascension of Ascending Node for plane $\rho$ [degrees]\\
$s_{\text{per plane}}$ & number of satellites in each plane of Walker-Star constellation \\
$s$ & satellite in-plane index of Walker-Star constellation \\
$S_{\rho,s}$ & Walker-Star satellite with plane and in-plane index $\rho,s$ \\
$M_{\rho,s}$ & mean anomaly of satellite $S_{\rho,s}$ in Walker-Star constellation [degrees]\\

$f_{\text{gap}}$ & mean gap time minimization objective function\\
$g_{\text{gaps}}$ & all gaps in contact windows \\
$\bar{g}_{\text{gaps}}$ & per-satellite mean gap time in contact windows\\
$n_{\text{gaps}}$ & count of all gaps in contact windows \\

\end{longtable*}}

%% file: 01_Introduction.tex
\section{Introduction}

The proliferation of satellite constellations in recent years is transforming global connectivity, planetary observation, and scientific exploration. Over the past decade, satellite deployment rates have accelerated markedly~\cite{eftimiades_implications2022, falle_one_2023, karacalioglu_impact2016}, driven by decreasing launch costs through ride-share missions~\cite{chen_concept_2024} and the declining cost of small satellite development. This rapid growth has increased the demand for reliable ground station infrastructure that can meet latency, cost, and capacity requirements~\cite{rasila_national_2024}. Selecting ground station locations poses a challenging decision for mission operators, as it is a multi-constrained problem driven by orbital alignment, access opportunities, cost, and site availability. Satisfying coverage, bandwidth, atmospheric conditions, infrastructure, and revisit requirements demands careful planning between satellite trajectories and ground segment assets, motivating the need for systematic location selection strategies.

Despite its operational importance, ground station network optimization remains relatively underexplored in the literature. In this work, we solve the core ground station planning problem of selecting up to a predefined number of ground station locations that optimize the performance of a given mission objective. We categorize the ground station planning problem into two distinct variants: free placement and fixed-site selection. In the free placement problem, candidate locations may vary continuously in longitude and latitude, while the fixed-site selection problem restricts choices to a discrete set of predefined locations. These two variants address either end of the spectrum of planning environments, ranging from fully flexible deployments for newly designed networks to configurations constrained by existing infrastructure. We note that free placement considers only geometric (longitude and latitude) optimization, excluding terrain suitability (slope, obstructions) or civil engineering constraints characteristic of "greenfield" deployments. We focus particularly on ground station placement optimization in the context of serving low Earth orbit (LEO) Earth observation (EO) satellite constellations, which introduces the challenge of satellites continuously moving with respect to the station locations. This is in contrast to geostationary satellites, which are generally in continuous view of their supporting ground station network.

The number of EO satellites in low Earth orbit has surged in recent years. As of 2024, there are 589 EO satellites in orbit, operated by 53 different commercial entities~\cite{wilkinson_environmental_2024}. These satellites support a broad range of applications, including precision agriculture~\cite{aragon_cubesats_2018, nguyen_monitoring_2020}, disaster monitoring for wildfire detection~\cite{chuvieco_satellite_2020}, and water resource management~\cite{sheffield2018satellite}. Due to the high-resolution sensors onboard, EO satellites produce significant volumes of imagery data. Most EO satellites produce over one terabyte (TB) of data per day~\cite{tao_transmitting_2023}, with some, such as NASA’s NISAR mission, delivering over 80 TBs daily~\cite{manavalan_nisar_2020}. These satellites typically operate in LEO between 500–650 km altitude, leading to short visibility windows with respect to individual ground stations~\cite{aragon_cubesats_2018}. Communication windows for LEO satellites typically last 3 to 10 minutes per access~\cite{vasisht_distributed_2020}. Collected data can be stored on board the spacecraft temporarily, but mission operators must maintain a sufficient downlink cadence to avoid buffer overflows, data loss, and data delivery delays that can severely impact operations~\cite{wang_infiltrating_2024}. Sub-optimal ground station placement can drastically reduce the number of these opportunities, further limiting communications capabilities and impacting overall mission performance.

Prior work has largely studied optimization of spacecraft operations and contact scheduling given existing ground networks, rather than addressing the placement of ground infrastructure itself. Mixed Integer Linear Programming (MILP) is a commonly used technique in this space, offering optimal solutions to complex scheduling constraints. For example, the Shaving Model~\cite{linares_mixed_2024} used MILP to dynamically adjust pass durations across a fixed network, reducing missed connections by 23\% over traditional integer programming approaches~\cite{cho_optimization-based_2018}. The China remote sensing satellite ground station applied both MILP and particle swarm optimization to break down multi-station relay tasks into tractable single-station sub-problems~\cite{tian_optimizing_2023}. For constellations with 50 or more satellites, decentralized approaches such as geometric neighborhood decomposition have been introduced to localize decision-making while preserving global coordination~\cite{zilberstein_decentralized_2024}. Conflict resolution has also received significant attention: Terran Orbital, for example, implements a tiered priority system to allocate antenna time based on mission criticality~\cite{fitzgibbon_simple_2023}, while others explore secure scheduling via quantum key distribution frameworks~\cite{maule_fair_2024}. Despite these advancements in scheduling and operational efficiency, nearly all prior work assumes fixed ground station locations. As a result, the question of whether ground infrastructure is optimally placed to meet mission requirements remains largely unexamined.

Ground station and gateway placement optimization literature commonly varies across three key dimensions: number of objectives, optimization techniques, and the problem formulation. Single-objective works predominate in fixed-site downselection, where meta-heuristics and heuristics select optimal subsets from predefined candidate pools. For instance, \citeauthor{guo2021gateway}~\cite{guo2021gateway} employ particle swarm optimization, a meta-heuristic, to balance traffic load and deployment cost across fixed gateway candidates, while \citeauthor{liu2019throughput}~\cite{liu2019throughput} use throughput evaluation heuristics for ground station planning from existing sites. Multi-objective formulations expand to grid-down selection and joint optimization challenges, often blending meta-heuristics with mathematical programming. \citeauthor{baeza2023gateway}~\cite{baeza2023gateway} integrate rain attenuation, visibility, elevation, and traffic in a multi-criteria grid-based framework for NGSO gateways; their follow-on work \cite{baeza2023multi} adds capacity diversity through similar grid downselection. \citeauthor{abe2024optimizing}~\cite{abe2024optimizing} pursue joint placement-routing with mixed-integer programming on weighted multi-objectives of flow and cost, while 
\citeauthor{chen2021multiple}~\cite{chen2021multiple} apply MILP heuristics to multi-gateway coverage in ISL-enabled constellations from candidate sets. 

Pure free-placement problem formulations, optimizations performed continuously over longitude and latitude without predefined sites, are less prevalent due to computational complexity. Architecture studies from \citeauthor{del2016architecting} \cite{del2016architecting,del_portillo_optimal_2017} leverage genetic algorithms for optical/EHF ground segment design, trading availability, cost, and capacity while minimizing cloud cover and latency. A larger body of prior work has focused on selecting the best subset of stations from a predefined pool of existing candidate sites, rather than exploring methods for free placement of entirely new stations. In the EO context,~\citeauthor{eddy_optimal_2024}~\cite{eddy_optimal_2024} formulate an integer programming (IP) based approach to identify data- and cost-optimal configurations using commercial Ground Station-as-a-Service (GSaaS) providers. For optical ground station (OGS) networks,~\citeauthor{fuchs_optimization_2017}~\cite{fuchs_optimization_2017} and others select the most effective subset of existing OGSs to support reliable space-to-ground optical communications~\cite{fuchs_ground_2015, fuchs_optimization_2017}. While candidate expansion and downselection suffice when infrastructure aligns with mission needs, terrain suitability, regulatory restrictions, power access, backhaul, and atmospheric factors often exclude optimal locations from the commercial pool. While pure continuous search methods address this gap, their computational cost hinders convergence for large networks; this necessitates scalable free placement optimization methods for greenfield networks supporting mission-scale LEO constellations.

Outside of the space domain, a substantial body of literature exists on free placement location optimization in operations research~\cite{cui_distribution_2023}, telecommunications~\cite{sachan_genetic_2016}, and urban management~\cite{ayati2025optimizing}. The optimization of warehouse, hospital, fire station, and cell tower locations to maximize coverage is a classic facility location problem, often addressed using meta-heuristic algorithms such as differential evolution~\cite{kim_robust_2020,wang_critical_2023}, genetic algorithms~\cite{sachan_genetic_2016, raisanen_comparison_2005}, particle swarm optimization~\cite{guner2008discrete}, and ant colony optimization~\cite{xu_localization_2022}. These approaches navigate large, complex solution spaces, enabling flexible placement while balancing competing objectives such as cost, coverage, and demand. In such problems, gradient-based methods are often unreliable, as gradients may not exist at all feasible points or may fail to guide the search toward a global optimum. Similar characteristics appear in free placement ground station optimization problem, where discontinuous, non-convex objectives complicate the search. Adapting meta-heuristic or trial-and-error methods from other domains is therefore a natural approach, as they can explore these complex spaces and achieve near-optimal solutions~\cite{crawford_near-optimal_2000, raisanen_comparison_2005, liu_optimising_2010}. Despite their effectiveness, even near-optimal solutions are computationally expensive in large search spaces, highlighting the need for more efficient methods~\cite{bagchi_near_2009, crawford_near-optimal_2000}.

This work conducts a comprehensive study on free placement ground station optimization for LEO Earth observation satellite constellations. We introduce SCORE (Sequential Cyclic Optimization via Refinement \& Evaluation), a framework for selecting ground station locations through iterative, cyclic coordinate search. Unlike traditional global optimization methods, SCORE decomposes the high-dimensional placement problem into a sequence of focused selection steps, enabling efficient exploration of large decision spaces. We evaluate SCORE’s performance through two comparative studies. First, we benchmark the algorithm against differential evolution, a free-placement metaheuristic method, where we demonstrate that SCORE achieves comparable or superior station placements while reducing the number of function evaluations needed for convergence by 5$\times$. Second, to assess the advantages of continuous coordinate free-placement optimization over fixed-site selection, we compare SCORE to a globally optimal IP formulation that selects sites specifically from KSAT and locations listed by the World Teleport Association. SCORE’s continuous free-placement optimization identifies placement configurations that outperform fixed-site methods by up to 15\%, highlighting the potential gains from flexible placement. Under operationally feasible latitude constraints, SCORE retains over 92\% of this gain while remaining within commercially viable deployment regions. We demonstrate these advantages in case studies involving the 5-satellite Capella constellation and the 34-satellite ICEYE constellation, and explore trade-offs between deploying additional ground stations versus expanding antenna capacity at current infrastructure sites.





%% file: 02_test_probFormulation.tex
\section{Problem Formulation}
\label{sec:ProblemFormulation}

The ground station placement problem is selecting a set of $n$ longitude–latitude $(\lambda, \phi)$ coordinate locations $\mathcal{L}$ for a given satellite constellation $\mathcal{S}$ that maximizes a mission-specific performance objective $f$. Ideally, the ground station placement problem could be posed as a scheduling optimization by predicting all possible satellite-to-ground contact opportunities over the entire mission duration, then identifying an optimal set of contact opportunities for this time period. This formulation would guarantee optimal performance over the entire mission duration, but due to the stochastic orbital perturbations experienced by LEO spacecraft, 
it is not possible to perform long-term orbital propagation with sufficient accuracy to predict all contact opportunities over multiple years. 

To address this challenge, we adopt a surrogate optimization inspired by~\citeauthor{eddy_optimal_2024}~\cite{eddy_optimal_2024} that focuses on performing ground station placement optimizations over shorter time periods, typically between 7 and 10 days, rather than the full multi-year-long mission duration. The cyclic nature of satellite orbits means that such a window captures multiple orbital cycles, producing a distribution of contact opportunities that is assumed to approximate the long-term contact opportunities of the constellation. While seasonal variations do affect the exact contact opportunities over long durations, the overall number and order over the shorter duration are assumed to be a representative surrogate. We perform a large-scale computational study in Appendix  \ref{app:SurrgOpt} to validate this assumption. The full mission duration is defined by the optimization start time $t_{\text{opt}}^{\text{start}}$ and end time $t_{\text{opt}}^{\text{end}}$, with total duration $T_{\text{opt}} = t_{\text{opt}}^{\text{end}} - t_{\text{opt}}^{\text{start}}$. For surrogate optimization, a shorter simulation window is used, defined by start and end times $t_{\text{sim}}^{\text{start}}$ and $t_{\text{sim}}^{\text{end}}$, respectively, resulting in a fixed one-week duration $T_{\text{sim}} = t_{\text{sim}}^{\text{end}} - t_{\text{sim}}^{\text{start}}$.

We start our problem formulation by defining the set of all coordinate points $\mathcal{G}$ on Earth's ellipsoidal surface, where
\begin{equation}
\mathcal{G} = \left\{ (\lambda, \phi) \in \mathbb{R}^2 \ \mid\ -180^\circ \leq \lambda \leq 180^\circ,\ -90^\circ < \phi < 90^\circ \right\} \cup  \left\{(0, -90),  (0, 90)\right\} 
\end{equation}
is the set of all possible longitude-latitude coordinate pairs ($\lambda_i,\phi_i$). A ground station network is then defined as
\begin{equation}
\mathcal{L} = \{(\lambda_1, \phi_1), \dots, (\lambda_n, \phi_n)\} ,\quad \mathcal{L}\subseteq\mathcal{G},\quad \lvert \mathcal{L}\rvert=n    
\end{equation}
where $\mathcal{L}$ is the set of the final $n$ selected site locations. Every station $L \in \mathcal{L}$ has a fixed data rate $L_{\text{dr}}$, representing the maximum number of bits that can be received per second at that site. In this work, $L_{\text{dr}}$ is assumed constant across all sites, though it can generally vary between stations. The network selection is optimized with respect to a set of satellites $\mathcal{S}$, where each spacecraft $S \in \mathcal{S}$ has associated design constants such as its orbital elements (two-line-elements, or TLEs, in this work) and fixed data rate $S_{\text{dr}}$. We study two problem formulations for selecting $\mathcal{L}$, free placement and fixed-site selection.

In the free placement problem, site locations $\mathcal{L}$ can be any longitude and latitude coordinate in $\mathcal{G}$. This formulation models cases where ground station network locations need to be selected from any continuously varying location. Additional constraints can reduce the search space of $\mathcal{G}$, such as limiting placement to land-based locations $\mathcal{T}$ where $\mathcal{T} \subseteq \mathcal{G}$, or further restricting to infrastructure-accessible zones $\mathcal{I} \subseteq \mathcal{T}$ defined by proximity to existing population centers and terrestrial infrastructure. In this work, free placement optimization operates over geometric coordinate locations and does not account for terrain suitability or civil engineering constraints in traditional ``greenfield'' site selection.

For the fixed-site selection problem, candidate ground station sites for the final ground network $\mathcal{L}$ are drawn from a discrete predefined coordinate list of ground stations $\mathcal{P}$. Each location $(p, P) \in \mathcal{P}$ has an associated binary decision variable $p \in \{0,1\}$ that is 1 if the location is selected and 0 otherwise. Location $P$ is a coordinate $(\lambda,\phi)$. The final set of selected ground station sites is then defined as 
\begin{equation}
    \mathcal{L} = \{P\mid(p, P) \in \mathcal{P}, p = 1\}
\end{equation}
the subset of all locations with $p = 1$. This fixed-site selection problem is more applicable in the GSaaS or teleport selection cases, where existing ground station facilities or foundational infrastructures already exist and the selection problem is restricted to choosing from these preexisting locations.

The selected networks are evaluated based on the quality of their possible contact windows $\mathcal{C}_{\mathcal{L},\mathcal{S}}$, which represent all opportunities for communication between the network $\mathcal{L}$ and spacecraft $\mathcal{S}$ during the simulation window $T_{\text{sim}}$. Since our goal is to estimate the achievable communication performance under realistic operational constraints, we formulate a scheduling problem over these contacts to approximate the network capacity achievable during the mission. Similar to $\mathcal{P}$, each contact $(c, C) \in \mathcal{C}_{\mathcal{L},\mathcal{S}}$ contains a binary decision variable $c\in\{0,1\}$ that indicates whether that contact has been selected for scheduling. Each contact $C$ has a variety of other constants, such as the contact start time and end time ($C^{\text{start}}$, $C^{\text{end}}$). The data rate for each contact $C_{\text{dr}}$ is set as the minimum of the satellite and ground station data rates 
\begin{equation}
\begin{aligned}
	C_{\text{dr}} = \min(L_{\text{dr}},S_{\text{dr}})
\end{aligned}
\label{eqn:contact_datarate}
\end{equation}
participating in the contact. The duration of the contact is taken as the difference between the contact end and start times $C_{\text{duration}} =  C^{\text{end}} - C^{\text{start}}$. 

\subsection{Objective Functions} 
\label{sec:objective_func}

Mission operators may prioritize different objectives when selecting a ground station network to support a satellite constellation. In general, these objectives can be expressed as selecting a set of sites that yields the most favorable contact windows $\mathcal{C}_{\mathcal{L,S}}$ between the spacecraft and the ground network, subject to the mission goals and constraints. We first define a general optimization formulation as
\begin{equation}
    \max_{\mathcal{L} \subset \mathcal{G}} f(\mathcal{L}, \mathcal{S})
\label{eqn:generalOpt}
\end{equation}
where the decision variable is a candidate set of ground stations $\mathcal{L}\subset \mathcal{G}$ and $f(\mathcal{L,S})$ is an objective function that evaluates the quality of the chosen network given the constellation $\mathcal{S}$.

This work focuses on designing ground station networks for high-data-volume EO constellations, with the primary goal of maximizing total downlinked data over the mission, independent of delivery latency. We call this the \textit{data downlink-maximization objective}. To evaluate this objective for a given candidate network $\mathcal{L}$, we first formulate a scheduling problem that determines which contacts between each spacecraft and the ground stations in $\mathcal{L}$ are feasible, subject to network and visibility constraints. Visibility is enforced through a minimum elevation angle requirement, such that a contact is feasible only if the satellite elevation exceeds $10^\circ$. Network constraints are further detailed in \Cref{sec:regularization,sec:constraintFixed}. Only contacts selected in the scheduling ($c = 1$) are included in computing the final objective. The total data downlinked from a network $\mathcal{L}$ is defined as
\begin{equation}
 f_{\text{data}}(\mathcal{L,S}) = \frac{T_{\text{opt}}}{T_{\text{sim}}}\sum_{L \in {\mathcal{L}}}\sum_{(c,C) \in {\mathcal{C}_L}}C_{\text{dr}}C_{\text{duration}} c
\label{eqn:obj_max_data}
\end{equation}
where $\mathcal{C}_L\subseteq \mathcal{C}_{\mathcal{L,S}}$ represents the set of contacts associated with station $L$. We weight the objective by $\frac{T_{\text{opt}}}{T_{\text{sim}}}$ to appropriately approximate the total data volume downlinked over the mission duration, rather than only the simulated time window. We also outline other objectives such as mean-gap minimization; details are provided in Appendix \ref{app:meangap}.

\subsection{Regularization: Reducing the Search Space}
\label{sec:regularization}
For the free placement problem, additional constraints need to be imposed on the search space $\mathcal{G}$ to prevent the selection of infeasible locations. We consider three constraints that define infeasible ground station placements: stations must be located on land; stations must maintain a minimum separation distance to prevent overlapping communication cones; and stations must be close to existing power and terrestrial communications infrastructure. However, if we incorporate every infeasible region as a hard constraint, we limit the optimization methods that can be used to solve the problem. Many efficient optimization techniques, such as gradient-based methods or global heuristics designed for unconstrained problems, are not naturally equipped to handle highly irregular, non-convex feasible regions, making it impossible or computationally expensive to enforce hard constraints directly. To address this, we incorporate these constraints as penalty terms subtracted from the objective function in \Cref{sec:objective_func}, discouraging violations without restricting the optimization method. Each penalty takes the form of a quadratic loss function
\begin{equation}
    Q(L) = (\| L - L^*\|)^2
    \label{eqn:penalty}
\end{equation}
where $L^*$ represents the nearest feasible ground station location that satisfies the constraint on $L$. When the constraint is satisfied, $\|L - L^*\| = 0$ and no penalty is applied. When violated, the penalty grows quadratically with the distance to the nearest feasible point. This provides a flexible tool for modeling ground station placement constraints, as any hard constraint can be incorporated into the objective function in this form.

We first apply this to the \textit{land placement constraint}. This constraint ensures that all ground stations can only be placed in feasible terrestrial locations $\mathcal{L} \subseteq\mathcal{T}$. Although this constraint focuses on excluding bodies of water, the definition of $\mathcal{T}$ can be modified to reduce the search space to exclude any infeasible station placement location. We transform this hard constraint into a penalty function like in \Cref{eqn:penalty} as
\begin{equation}
    Q_\mathcal{T}(\mathcal{L}) = \sum_{L\in\mathcal{L}}(\| L- L_{\mathcal{T}}\|)^2
    \label{eqn:penalty_land}
\end{equation}
where we calculate the geodesic distance between station $L$ and $L_{\mathcal{T}}$, which represents the closest land mass coordinate to station $L$ in $\mathcal{T}$. If station $L$ is in $\mathcal{T}$, the distance between $L - L_\mathcal{T}$ is zero. This penalty can be subtracted from our objective functions in \Cref{sec:objective_func}, to penalize whenever a given ground station is not within the feasible terrestrial locations. 
The penalty grows quadratically the further station $L$ is from $\mathcal{T}$. With this formulation of the quadratic loss penalty, any geographic region becomes heavily penalized in the search space, effectively discouraging its selection in the free placement ground station optimization problem.

Next, we consider the \textit{station proximity constraint}, which ensures a minimum distance $d_{\min}$ is maintained between all ground stations so that network location redundancy is minimized and closely spaced ground stations do not experience interference limitations. In practice, this restriction may be relaxed through techniques such as polarization diversity or frequency allocation at high-demand ground sites, enabling multiple antennas to operate in close proximity. Incorporating such capabilities would require extending the formulation to jointly model spatial placement and communication resource allocation, which is beyond the scope of this work but represents an interesting direction for future research. 

Given a set of selected candidate locations $\mathcal{L}$, $d_{L_i,L_j}$ is denoted as the distance between locations $L_i$ and $L_j$. for all locations $L_i, L_j \in\mathcal{L}$. The value of $d_{\min}$ is a predefined constant of the minimum allowable distance between any two stations. The constraint is formulated as
\begin{equation}
\begin{aligned}
   	&d_{L_i,L_j} > d_{\min} \hspace{1em}\forall \; L_i, L_j \in {\mathcal{L}} \text{ where } L_i \neq L_j \\
\end{aligned}
\label{eqn:station_proximity_constraint}
\end{equation}
which ensures that any pair of candidate ground station locations $L_i,L_j \in \mathcal{L}$, maintains a minimum separation distance of at least $d_{\min}$. To regularize this constraint, we frame \Cref{eqn:station_proximity_constraint} again in the form of the general penalty function from \Cref{eqn:penalty}
\begin{equation}
    Q_{\text{proximity}}(\mathcal{L}) = \sum_{i=0}^{n-1} \sum_{j = i+1}^{n}\max(0, d_{\min}-d_{L_i, L_j})^2\label{eqn:penalty_proximity}
\end{equation}
where $Q_{\text{proximity}}(\mathcal{L})$ is calculated between every combination of $L_i, L_j \in \mathcal{L}$. The penalties associated with each $L_i, L_j$ pair in \Cref{eqn:penalty_proximity} can again be subtracted from the objective function $f(\mathcal{L,S})$ to apply these constraints.

Finally, we consider an \textit{infrastructure proximity constraint}, which encourages ground station placement near existing population centers and 
terrestrial infrastructure. This constraint is motivated by the practical requirement that ground stations require access to reliable power and fiber backhaul for high-throughput data downlink operations. We define a set of feasible infrastructure zones $\mathcal{I} \subseteq \mathcal{G}$ as the union of circular regions of radius $r_{\text{infra}}$ centered on known population centers, such that
\begin{equation}
    \mathcal{I} = \bigcup_{k} \{ (\lambda, \phi) \in \mathcal{G} \mid 
    d((\lambda,\phi), (\lambda_k, \phi_k)) \leq r_{\text{infra}} \}
    \label{eqn:infra}
\end{equation}
where $(\lambda_k, \phi_k)$ are the coordinates of known population centers and $r_{\text{infra}}$ is the maximum allowable distance from infrastructure, set to 50~km in this work. Similar to the land placement constraint, we 
transform this into a penalty function as
\begin{equation}
    Q_{\mathcal{I}}(\mathcal{L}) = \sum_{L \in \mathcal{L}} 
    (\| L - L_{\mathcal{I}} \|)^2
    \label{eqn:penalty_infra}
\end{equation}
where $L_{\mathcal{I}}$ represents the closest point in $\mathcal{I}$ to station $L$. The full regularized optimization problem from \Cref{eqn:generalOpt} can now be written as
\begin{equation}
    \max_{\mathcal{L}} f(\mathcal{L,S}) - Q_\mathcal{T}(\mathcal{L}) - 
    Q_{\text{proximity}}(\mathcal{L}) - Q_{\mathcal{I}}(\mathcal{L})
\label{eqn:FullOpt}
\end{equation}
with the regularization terms of the land placement, proximity between stations, and proximity to infrastructure penalty. For the fixed-site selection case, these penalties are set to zero as all locations in $\mathcal{P}$ are already pre-filtered to guarantee that they do not violate these conditions. With the above formulation, the free-placement problem is fully specified and can be solved using existing optimization techniques such as gradient-based search or evolutionary algorithms. However, for the fixed-site selection variation, additional work is needed to express the formulation as an integer program that explicitly models discrete decisions.

\subsection{Constraints and Fixed-Site Integer Programming Formulation}
\label{sec:constraintFixed}
After selecting objective and regularization terms, constraints are introduced to model system, design, and operational limitations. These can be grouped into two categories, contact exclusion constraints, which prevent infeasible overlaps in scheduling, and selection constraints, which enforce consistency between contacts, sites, and network size. These are necessary to properly compute the maximum feasible number of contacts that can be taken and data downlinked given a set of predicted contacts. In the free placement case, contact exclusion constraints are applied during evaluation after candidate networks are selected, ensuring fairness when comparing to fixed-site results. In fixed-site selection, both contact exclusion and selection constraints are encoded directly into an integer program. This formulation provides a unified optimization framework: free placement problem formulations are optimized through penalty-regularized search in \Cref{eqn:FullOpt}, while fixed-site selection leverages integer programming formulation solvers with exact optimality guarantees.  Together, they enable consistent evaluation of placement strategies under different operational contexts.

\subsubsection{Contact Exclusion Constraints}
Contact exclusion constraints model the limitations of the spacecraft or ground station site's ability to communicate with different assets. These constraints reflect the single-antenna limitation present either on satellites or at ground stations, ensuring that at any given time, each satellite communicates with at most one ground station and vice versa. To build the \textit{satellite contact exclusion constraint}, all contacts for satellite $S$, denoted as $\mathcal{C}_S$, must be examined to identify any simultaneous contacts from multiple sites within candidate network $\mathcal{L}$. For computational efficiency, the contacts can be sorted in ascending time-order to limit checks to temporally nearby contacts, though this is not required for the constraint itself. This constraint can be expressed as
\begin{equation}
\begin{aligned}
	& c_i + c_j \leq 1 \; \hspace{1em} \forall \; S \in {\mathcal{S}}, i,j \in \{1,\ldots,\lvert{\mathcal{C}}_S\rvert\}, j > i \; \text{s.t.} \\
   	& \hspace{1em} C^{\text{start}}_{i} \leq C^{\text{end}}_{j} \\
   	& \hspace{1em} C^{\text{start}}_{j} \leq C^{\text{end}}_{i} \\
\end{aligned}
\label{eqn:constraint_contact_exclusion}
\end{equation}
where only one contact in every pair of overlapping contacts for a satellite is selected, leading to each satellite communicating with only one ground station at a time. The inverse of the satellite contact exclusion constraint can also be formulated to reflect ground station infrastructure limitations, where each station is restricted to serving only one satellite simultaneously. For the \textit{station contact exclusion constraint}, the constraint is outlined as 
\begin{equation}
\begin{aligned}
   	& c_i + c_j \leq 1 \; \hspace{1em} \forall \; L \in {\mathcal{L}}, i,j \in \{1,\ldots,\lvert\mathcal{C}_L\rvert\}, j > i \; \text{s.t.} \\
   	& \hspace{1em} C^{\text{start}}_{i} \leq C^{\text{end}}_{j} \\
   	& \hspace{1em} C^{\text{start}}_{j} \leq C^{\text{end}}_{i} \\
\end{aligned}
\label{eqn:constraint_station_contact_exclusion}
\end{equation}
where each location is set to ensure communication can only occur with one satellite at any time. All contacts in $\mathcal{C}_L$ are organized in time-order sequence.

\subsubsection{Fixed-Site Integer Programming Selection Constraints}

The fixed-site ground station selection problem is particularly well suited for formulation as an integer program, as both the predefined ground station selection list $\mathcal{P}$ and set of contacts $\mathcal{C}(\mathcal{L},\mathcal{S})$ have associated binary decision variables indicating whether the contact $(c,C)$ or station site $(p,P)$ were selected. By coupling site-selection variables $p$ with contact-scheduling variables $c$, the IP formulation enables simultaneous optimization of station locations and communication schedules. This approach allows us to leverage established solver libraries such as Gurobi \cite{gurobi} or COIN-OR \cite{lougee2003common}, which provide fast, efficient solutions to IP problems along with optimality certificates that indicate whether a solution is globally optimal, suboptimal, or infeasible. We follow existing integer programming selection formulations from prior works~\citep{eddy_optimal_2024,kim2026scalable}, which solve the fixed-site ground station selection problem for GSaaS providers.

To properly characterize the fixed-site selection ground station optimization problem using an IP, we consider the data downlink maximization objective defined in \Cref{sec:objective_func} and apply both contact exclusion constraints. We also introduce several additional IP selection constraints to ensure consistency between contact- and site-level decisions, enforcing network design requirements. The first additional constraint is to ensure that if a contact $c$ is selected, the decision variables of the location $p$ are also selected. No contacts should be scheduled at a location unless the location is activated. This is denoted as
\begin{equation}
\begin{aligned}
   	\sum_{(c, C) \in {\mathcal{C}}_P}c \leq \lvert{\mathcal{C}}_P\rvert p, \; \forall \; (p,P) \in {\mathcal{P}}
 \end{aligned}
\label{eqn:constraint_location_selection}
\end{equation}
where $\mathcal{C}_P$ represents the set of contacts from site $P$. Second, we introduce a \textit{network size constraint} to enforce the size constraint $n$ on the ground station network 
\begin{equation}
\begin{aligned}
    \sum_{i} p_i=n,   \forall \; (p,P) \in {\mathcal{P}}
\end{aligned}
\label{eqn:constraint_network_size}
\end{equation}
to ensure we select ground station networks with the intended size. We also introduce a \textit{minimum contact duration constraint} to ensure that only contacts with a duration greater than a specified $t_{\text{min}}$ are considered for planning. The constraint is defined as 
\begin{equation}
\begin{aligned}
   	& c = 0 \; \text{if} \; C^{\text{end}} - C^{\text{start}} < t_{\text{min}}, \forall \; (c,C) \in {\mathcal{C}}
\end{aligned}
\label{eqn:constraint_min_duration}
\end{equation}
which eliminates short-duration contacts that might not be operationally useful.

%% file: 03_Methods_FreeSelection.tex
\section{Methodology}

We outline methodologies for addressing the free placement ground station problem described in \Cref{sec:ProblemFormulation}. To navigate the large, high-dimensional search space, we employ two gradient-free optimization methods. First, we introduce Sequential Cyclic Optimization via Refinement \& Evaluation (SCORE), which combines sequential cyclic search with single-step, gradient-free optimization methods; in this work, we demonstrate its application using Nelder-Mead~\cite{nelder_simplex_1965} and Powell~\cite{powell1964efficient}. As SCORE is optimizer-agnostic, other gradient-free methods can be easily substituted for the single-step optimizer without changing the framework. Nelder-Mead was selected due to its simplicity, broad applicability in gradient-free settings, and well-understood convergence behavior in low-dimensional spaces. It is particularly suitable for local refinement, which aligns naturally with SCORE’s sequential cyclic updates. Powell was selected for its robustness to noisy or irregular objectives and its efficient line-search approach, providing a complementary local optimization strategy to Nelder-Mead within SCORE. Second, we review differential evolution (DE), a widely used evolutionary algorithm for location selection~\cite{kim_robust_2020,wang_critical_2023}. DE is a population-based global optimizer that serves as a strong baseline for comparison. While it can achieve high-quality solutions, it requires substantially more objective evaluations than SCORE, especially as the number of stations grows.

\input{SCORE}
\subsection{\textbf{Sequential Cyclic Optimization via Refinement \& Evaluation (SCORE)}}
To address the computationally costly nature of one-step global optimization methods, we introduce SCORE, Sequential Cyclic Optimization via Refinement \& Evaluation. SCORE decomposes the full optimization into a series of smaller problems and proceeds in two phases. (1) \textbf{Sequential coordinate selection}: beginning with an empty set, SCORE sequentially adds ground stations to $\mathcal{L}$, selecting at each step the longitude–latitude pair that maximizes the objective function given the current network configuration. This repeats until network $\mathcal{L}$ reaches the prescribed network size $n$. (2) \textbf{Sequential cyclic refinement}: with $n$ stations, the algorithm iteratively re-optimizes one station at a time while holding the others fixed, cycling through all stations to refine their positions and improve the objective, until a stopping criterion is met (e.g., negligible improvement or a budget on iterations). By reusing intermediate contact evaluations $\mathcal{C}_{\mathcal{L,S}}$ from modifying only one station $L\in\mathcal{L}$ each time, SCORE reduces the number of contact evaluations of the entire network needed during the search process. This coordinate structure-aware strategy enables SCORE to retain solution quality while significantly lowering computational cost, making it more practical for large-scale or design-iteration-heavy scenarios. We define the full framework in \Cref{alg:score}.

In the first phase, sequential coordinate selection, the algorithm constructs the ground station set $\mathcal{L}$ incrementally. Starting from an empty set, it repeatedly adds a station location $L \in \mathcal{G}$ that yields the largest improvement to the objective relative to the stations already chosen. This new location is added to $\mathcal{L}$, and the process continues until the desired number of stations $n$ is reached. Each added location $L$ is locally optimal based on the existing set $\mathcal{L}$, and the final combination of stations approaches near-optimal configurations at a fraction of the computation cost. SCORE's second phase, sequential cyclic refinement, performs further fine-tuning of the combined set of stations. Here, we refine the station set through iterative updates. For each station $L \in \mathcal{L}$, the algorithm temporarily removes $L$ to form a reduced set $\mathcal{L}' = \mathcal{L} \setminus \{L\}$, and re-optimizes that location using the same optimizer. The updated station $L'$ is then added back to form a new candidate set $\mathcal{L}' \cup \{L'\}$. The objective function values of the original set $\mathcal{L}$ and the modified set $\mathcal{L}' \cup  \{L'\}$ are compared, and if the new set improves performance, it replaces the current one. This process is repeated for each station $L \in \mathcal{L}$, and up to $E_{max}$ total cycles. The optimization terminates early if a full pass through all stations results in no updates, indicating that the current set $\mathcal{L}$ has converged to a locally optimal solution. Any gradient-free optimizer can be used for the \textsc{Optimizer} noted for each coordinate selection in \Cref{alg:score}. For this paper, we employ the Nelder-Mead simplex search algorithm \cite{nelder_simplex_1965} as well as Powell's conjugate direction method~\cite{powell1964efficient}.

\input{nelderMead}

\subsubsection{Nelder-Mead}
Similar to evolutionary algorithms, Nelder-Mead is a gradient-free direct search method with applications ranging from engineering \cite{ijaz2016fractional}, chemistry
\cite{bezerra_simplex_2016}, control theory \cite{nobahari_simplex_2016}, and computer science \cite{nobahari_simplex_2016}.
The simplex, a polytope in $k$ dimensional space with $k+1$ points \cite{mohsin_modified_2021}, is at the core of understanding how optimization is performed under Nelder-Mead. In the free-placement ground station problem, each vertex of the Nelder–Mead simplex corresponds to a candidate configuration of ground stations, denoted $\{ L_1, \dots, L_{k+1} \}$, where each $L_i$ in this simplex encodes the longitude and latitude of station $i$. Thus, a simplex $\{L_1, \dots, L_{k+1}\}$ represents $k+1$ competing configurations of station placements in the search space. At each iteration, the simplex vertices are ordered by their objective values, and the worst point is reflected across the centroid $L_{\text{cen}}$ of the remaining $k$ points to explore a potentially better location. The behavior of this transformation is governed by four parameters: the reflection coefficient $\alpha > 0$, expansion coefficient $\gamma > 1$, contraction coefficient $\rho \in (0,1)$, and shrinkage coefficient $\sigma \in (0,1)$. If the reflection improves the objective sufficiently, the new point is accepted; otherwise, the algorithm attempts an expansion (if the reflection is very good) or a contraction. Two types of contraction exist: outside contraction, when the reflection is moderately good, and inside contraction, when the reflection is worse than the current worst. If neither contraction yields improvement, the algorithm shrinks the entire simplex toward the best point, scaled by $\sigma$. This iterative process continues until convergence or a maximum number of iterations $Z_{\max}$ is reached. The best vertex is then returned as the approximate solution. We provide the full pseudocode of Nelder-Mead in \Cref{alg:nelder}.

We make two implementation adjustments to the Nelder-Mead algorithm to appropriately adjust the method for use in the ground station placement problem. 
First, rather than optimizing in a two-dimensional longitude–latitude space $(k=2)$, we project candidate locations onto the three-dimensional unit sphere and perform the optimization in that domain $(k=3)$. This requires a simplex with $k+1=4$ vertices (a tetrahedron) instead of a triangle. Operating directly in three dimensions avoids degeneracies at the poles that arise in the longitude–latitude system, where all longitudes collapse to a single point at $\phi=\pm90\degree$. Second, we select randomized starting simplexes on the unit coordinate sphere with the largest possible simplex area. 
Specifically, we generate 100 random feasible points on the unit sphere and evaluate all 
$(k+1)$-point combinations to form candidate simplexes. The simplex with the largest volume is selected, ensuring a well-scaled and well-distributed starting configuration. Without such a spread, Nelder-Mead can become trapped in poor local minima due to insufficient coverage of the solution space.
\subsubsection{Powell's Method}

Powell's conjugate direction method \cite{powell1964efficient} is a derivative-free direct search optimizer that minimizes a function of $m$ variables through successive one-dimensional line minimizations. Similar to Nelder-Mead, Powell is a gradient-free method that only requires function evaluations. The method maintains a set of $m$ search directions ${\mathbf{r}_1, \dots, \mathbf{r}_m}$, initialized to the standard basis vectors ${\mathbf{e}_1, \dots, \mathbf{e}_m}$. At each outer iteration, the algorithm sequentially minimizes $f$ along each direction in turn, updating the current point $\mathbf{w}$ after each line minimization. The net displacement $\mathbf{v} = \mathbf{w} - \mathbf{p}$ accumulated across all $m$ line searches is then used as a final line minimization step, exploiting the aggregate direction of improvement across the full cycle. The key innovation of Powell's method is its direction update rule, which progressively builds a set of mutually conjugate directions. After each full cycle, the oldest direction $\mathbf{r}_1$ is discarded, the remaining directions are shifted, and the normalized displacement $\frac{\mathbf{v}}{|\mathbf{v}|}$ is appended as the new direction $\mathbf{r}_m$. For a strictly quadratic objective, this procedure guarantees that after $m$ complete cycles, all search directions are mutually conjugate with respect to the Hessian, and the exact minimum is found \cite{powell1964efficient}. For non-quadratic objectives such as $f$ in this work, the method provides fast practical convergence without requiring derivatives. The algorithm terminates when the displacement norm $|\mathbf{v}|$ falls below a tolerance $\varepsilon_{\text{pow}}$ or a maximum number of iterations $Z_{\max}$ is reached, returning the best point $\mathbf{w}$ found. We provide the full pseudocode in \Cref{alg:powell}.

\input{powell}



\subsection{Differential Evolution}

Differential evolution~\cite{storn_differential_1997} is a robust, non-gradient-based meta-heuristic capable of efficiently exploring complex, high-dimensional solution spaces, making it suitable for a wide range of global optimization problems. The method is similar to evolutionary approaches such as genetic algorithms, but DE uses a population of real-valued vectors rather than encoded genotypes. With a real-valued vector population, the mutation and recombination steps can occur directly in continuous solution spaces without requiring encoding or decoding. New vectors in the population are generated from combinations of existing candidates, and the DE algorithm always retains the best-scoring solution. This black-box approach does not require gradient information to guide optimization.

To apply DE to the free placement ground station problem, the ground station candidate list $\mathcal{L}$ is decomposed into a one-dimensional vector of ordered longitude and latitude coordinates, denoted as $\mathbf{x}$. The algorithm maintains a population of such candidate solutions $\mathcal{X}=\{\mathbf{x}_i\}^{NP}_{i=1}$, of population size $NP$, balancing exploration with computational cost. New candidate vectors are generated using the mutation scale factor $F \in [0,2]$, which sets the magnitude of modifications to each vector, governing the trade-off between global exploration and local refinement at each iteration. Components of these mutant vectors may replace the current candidates according to the crossover rate $CR \in [0,1]$, where $CR=1$ means every component of $\mathbf{x}$ is replaced, and $CR=0$ means no replacements occur. The process continues for a maximum number of generations $G_{\max}$, providing a practical stopping criterion. DE requires access to the objective function $f$, the dimensionality $D$ of $\mathbf{x}$ (equal to $2n$, as each of the $n$ ground stations is represented by a longitude and latitude coordinate), and the bounds for each dimension in $\mathbf{x}$. The full pseudocode is provided in \Cref{alg:differentialEvolution}.

\input{DiffEvolution}

The general DE outline consists of performing a mutation, crossover, and selection process for each vector $\mathbf{x}_i$ in $\mathcal{X}$. The mutation is computed using three distinct vectors in the maintained population, separate from the vector $\mathbf{x}_i$ being modified. The scaling factor of this mutation is dictated by $F$. Each dimension of vector $\mathbf{x}_i$ then undergoes the crossover step, where each index either keeps their original value, or crosses over with the mutated vector value based on the crossover rate $CR$. A random vector index $d \in 1, \dots ,D$ is selected beforehand to ensure that in the rare case that all indexes randomly are not selected to cross over, the index $d$ in vector $\mathbf{x}_i$ is ensured to have a mutation. The mutated vector is saved in $\mathbf{u}_i$. Finally, the algorithm selects between the original vector $\mathbf{x}_i$ and the modified vector $\mathbf{u}_i$, selecting the vector with the lower objective value. If the new vector has a more optimal objective function value, this vector replaces the existing vector $\mathbf{x}_i$ in the population. After the solution reaches convergence, or the maximum number of generations $G_{\text{max}}$ is completed, the candidate vector $\mathbf{x}_i$ with the lowest objective value is returned.

Although DE's non-gradient approach allows for the search of the discontinuous search space of ground station placement optimization, the method's reliance on a large number of function evaluations makes the method computationally expensive. For problems where each computation evaluation is lightweight, this overhead may be acceptable. However, in our case, where every candidate configuration $\mathbf{x}_i$, or ground station $L$ requires recomputing contacts $\mathcal{C}_{L,\mathcal{S}}$ for evaluation, the computational cost quickly becomes a bottleneck. In addition, the number of candidate solutions needed to be evaluated grows with the size of the ground station network $\mathcal{L}$ needed to be optimized, which limits DE's scalability for larger network design problems. SCORE addresses these computational challenges, as a more efficient, gradient-free alternative tailored to the structure of the ground station placement problem.

%% file: SCORE.tex
\newcommand{\code}[1]{\texttt{#1}}
\begin{algorithm}[bp!]
\caption{SCORE: Sequential Cyclic Optimization via Refinement \& Evaluation}\label{alg:score}
\begin{algorithmic}[1]

\Function {\textsc{SCORE}}{$\mathcal{S}$, $f$, $n$, $E_{\text{max}}$} \Comment{\textcolor{gray} {$n$: desired \# of ground stations, $E_{\text{max}}$: maximum \# cyclic evaluations}}
\State \textbf{Initialize} $\mathcal{L} \gets \emptyset$, evaluation count $e \gets 0$
\State \textcolor{lightgray}{(1) Initial Sequential Coordinate Selection}
\While{length of $\mathcal{L} < n$}
\State $L\leftarrow $ \textsc{Optimizer($\mathcal{L}$, $\mathcal{S}$, $f$)} \Comment{\textcolor{gray}{Optimizer can be any gradient free method}}
\State $\mathcal{L} \gets \mathcal{L} \cup \{L\}$ \Comment{\textcolor{gray}{Add $L$ to set of ground station locations $\mathcal{L}$}}
\EndWhile

\State \textcolor{lightgray}{(2) Sequential Cyclic Refinement}
\While{$e < E_{\text{max}}$}
\State \textbf{Store} copy of $\mathcal{L}$ as $\mathcal{L}_{1}$
\For{$L \in \mathcal{L}$}
\State $\mathcal{L}' \gets \mathcal{L} \setminus \{L\}$ \Comment{\textcolor{gray}{New set $\mathcal{L}'$ without current station location $L$}}
\State $L' \leftarrow $ \textsc{Optimizer($\mathcal{L}'$, $\mathcal{S}$, $f$)}  \Comment{\textcolor{gray}{Select new station $L'$}}
\State $\mathcal{L}' \gets \mathcal{L}' \cup \{L'\}$   \Comment{\textcolor{gray}{New set $\mathcal{L}'$ with current station location $L$}}
\If{$f(\mathcal{L}'$, $\mathcal{S})$ yields better objective value  than $f(\mathcal{L}$, $\mathcal{S})$}
\State $\mathcal{L} \leftarrow \mathcal{L}'$
\EndIf
\EndFor
\If{$\mathcal{L}_{1} = \mathcal{L}$ }\Comment{\textcolor{gray}{No changes were made, current sequence of stations are optimal}}
\State \textbf{break}
\EndIf
\State $e = e +1$
\EndWhile
\State \textbf{return} $\mathcal{L}$
\EndFunction

\end{algorithmic}
\end{algorithm}

%% file: nelderMead.tex
\begin{algorithm}[bp!]
\caption{Nelder-Mead (NM)}\label{alg:nelder}
\begin{algorithmic}[1]
\Function{NelderMead}{$f, \alpha, \gamma, \rho, \sigma, Z_{\max}$,$\mathcal{S}$}
\State Initialize simplex $\{L_1,\dots,L_{k+1}\}$ randomly; set $z \gets 0$
\While{$z < Z_{\max}$ and stopping condition not met}
    \State \textbf{Sort} simplex points $\{L_1, \dots, L_{k+1}\}$ so that $f(L_1,\mathcal{S}) \leq \dots \leq f(L_{k+1},\mathcal{S})$
    \State \textbf{Compute} centroid $L_{\text{cen}}$ of best $k$ points (excluding worst)
    
    \State $L_r \gets L_{\text{cen}} + \alpha(L_{\text{cen}} - L_{k+1})$
    \If{$f(L_1) \leq f(L_r) < f(L_k)$} 
        \State Replace $L_{k+1}$ with $L_r$  \Comment{\textcolor{gray}{(1) Reflection:}}
    \ElsIf{$f(L_r) < f(L_1)$} 
        \State $L_e \gets L_{\text{cen}} + \gamma(L_r - L_{\text{cen}})$
        \If{$f(L_e) < f(L_r)$}  \Comment{\textcolor{gray}{(2) Expansion:}}
            \State Replace $L_{k+1}$ with $L_e$ 
        \Else
            \State Replace $L_{k+1}$ with $L_r$
        \EndIf

    \ElsIf{$f(L_r) \geq f(L_n)$}
        \If{$f(L_r) < f(L_{k+1})$}
            \State $L_c \gets L_{\text{cen}} + \rho(L_r - L_{\text{cen}})$ \Comment{\textcolor{gray}{(3) Outside contraction}}
        \Else
            \State $L_c \gets L_{\text{cen}} - \rho(L_{\text{cen}} - L_{k+1})$ \Comment{\textcolor{gray}{(3) Inside contraction}}
        \EndIf
        \If{$f(L_c) < f(L_{k+1})$}
            \State Replace $L_{k+1}$ with $L_c$
        \Else
            \For{$i = 2$ to $k+1$ of $\{L_1, \dots, L_{k+1}\}$} \Comment{\textcolor{gray}{(4) Shrink:}}
    \State $L_i \gets L_1 + \sigma (L_i - L_1)$
\EndFor
        \EndIf
    \EndIf
    \State $z \gets z + 1$
\EndWhile
\State \Return best $L \in \{L_1, \dots, L_{k+1}\}$ in simplex such that $f(L,\mathcal{S})$ is minimized
\EndFunction
\end{algorithmic}
\end{algorithm}

%% file: powell.tex
\begin{algorithm}[tbp!]
\caption{Powell's Conjugate Direction Method}\label{alg:powell}
\begin{algorithmic}[1]
\Function{Powell}{$f, Z_{\max}, \varepsilon_{\text{pow}}$, $\mathcal{S}$}
    \State Initialize starting point $\mathbf{w}$; set search directions $\mathbf{r}_i \gets \mathbf{e}_i$ for $i = 1, \dots, m$; set $z \gets 0$
    \While{$z < Z_{\max}$ and stopping condition not met}
        \State $\mathbf{p} \gets \mathbf{w}$ \Comment{\textcolor{gray}{Save starting point for this cycle}}
        \For{$i = 1$ to $m$}
            \State $t^* \gets \arg\min_{t}\, f(\mathbf{w} + t\,\mathbf{r}_i,\, \mathcal{S})$ \Comment{\textcolor{gray}{Line minimization along $\mathbf{r}_i$}}
            \State $\mathbf{w} \gets \mathbf{w} + t^*\,\mathbf{r}_i$
        \EndFor
        \State $\mathbf{v} \gets \mathbf{w} - \mathbf{p}$ \Comment{\textcolor{gray}{Net displacement (new conjugate direction candidate)}}
        \If{$\|\mathbf{v}\| < \varepsilon_{\text{pow}}$}
            \State \textbf{break} \Comment{\textcolor{gray}{Convergence}}
        \EndIf
        \State $t^* \gets \arg\min_{t}\, f(\mathbf{w} + t\,\mathbf{v},\, \mathcal{S})$ \Comment{\textcolor{gray}{Line minimization along $\mathbf{v}$}}
        \State $\mathbf{w} \gets \mathbf{w} + t^*\,\mathbf{v}$
        \State \textbf{Update} directions: $\mathbf{r}_i \gets \mathbf{r}_{i+1}$ for $i = 1, \dots, m-1$; $\mathbf{r}_m \gets \mathbf{v}/\|\mathbf{v}\|$ \Comment{\textcolor{gray}{Drop $\mathbf{r}_1$, append $\mathbf{v}$}}
        \State $z \gets z + 1$
    \EndWhile
    \State \Return $\mathbf{w}$ such that $f(\mathbf{w}, \mathcal{S})$ is minimized
\EndFunction
\end{algorithmic}
\end{algorithm}

%% file: DiffEvolution.tex
\begin{algorithm}[htbp!]
\caption{Differential Evolution (DE)}\label{alg:differentialEvolution}
\begin{algorithmic}[1]

\Function {\textsc{DifferentialEvolution}}{$f$, $NP$, $F$, $CR$, $G_{\text{max}}$, $D$, $\mathcal{S}$} 
\State \textbf{Initialize} population $\mathcal{X}=\{\mathbf{x}_i\}_{i=1}^{NP}$ of dimensionality $D$ randomly within bounds
\State \textbf{Initialize} generation count $g \gets 0$
\While{$g < G_{\text{max}}$ and solution not converged}
    \For{each base vector $\mathbf{x}_i$ in population}
        \State \textcolor{lightgray}{(1) \textbf{Mutation:}}
        \State Randomly select 3 distinct vectors such that $\mathbf{x}_1 \neq \mathbf{x}_2 \neq \mathbf{x}_3 \neq \mathbf{x}_i$
        \State $\mathbf{v}_i = \mathbf{x}_{1} + F \cdot (\mathbf{x}_{2} - \mathbf{x}_{3})$
    
        \State \textcolor{lightgray}{(2) \textbf{Crossover:}}
        \State Pick a random vector index $d \in {1,…,D}$
        \For{$j = 1, \dots,d$}
            \State Generate $\text{rand}_j \sim U(0,1)$
            \If{$\text{rand}_j < CR$ \textbf{or} $j = d$}
                \State $\mathbf{u}_{i,j} = \mathbf{v}_{i,j}$
            \Else
                \State $\mathbf{u}_{i,j} = \mathbf{x}_{i,j}$
            \EndIf
        \EndFor
        \State \textcolor{lightgray}{(3) \textbf{Selection:}}
        \If{$f(\mathbf{u}_i,\mathcal{S}) \leq f(\mathbf{x}_i,\mathcal{S})$}
            \State $\mathbf{x}_i \gets \mathbf{u}_i$
        \EndIf
    \EndFor
    \State \textbf{Update} generation count: $g \gets g + 1$
\EndWhile
\State \textbf{return} best $\mathbf{x}_i$ in population such that $f(\mathbf{x}_i, \mathcal{S})$ is minimized
\EndFunction
\end{algorithmic}
\end{algorithm}

%% file: 05_ExperimentalSetup.tex
\section{Experimental Setup}

Before discussing the performance of SCORE against other ground station optimization methods, we outline our experimental simulation parameters for reproducibility. For all experiments, we used the simulation parameters outlined in \Cref{tab:sim_params} and propagated each spacecraft's dynamics using the SGP4 propagator \cite{vallado_revisiting_2006}.
Following the surrogate optimization framework in \Cref{sec:ProblemFormulation}, we restricted the simulation to the first 7 days ($T_{sim}$), which was scaled by $\frac{T_{opt}}{T_{sim}}$ to approximate the full mission duration $T_{opt}$. We applied a $10^\circ$ elevation mask at all ground station locations when calculating contact opportunities. In the current formulation, we assume a fixed data rate for all ground stations and satellites throughout each contact window, independent of elevation angle, and do not model frequency band assignments. All simulations were run on a dual-socket workstation with a total of 28 physical cores (56 threads), using 2.6 GHz Intel Xeon E5-2690 v4 processors and 128 GB of RAM.

\input{parameters}

\subsection{Satellite Constellations}

In our evaluations, we ran experiments on both synthetic constellations and real satellites, with orbits defined by Two-Line Element (TLE) sets. We used synthetic constellations specifically in parameter sweeps that explored the impact of varying constellation sizes. By generating synthetic constellations, we could precisely adjust simulation environments to measure how increasing constellation size affected the optimizer’s speed to reach convergence. For experiments that did not involve varying constellation sizes, we used existing EO satellite constellations, with TLEs providing orbital data for active satellites. In practice, satellite operators select ground stations from a limited set of locations offered by GSaaS providers to establish a supporting ground network. In this context, we compared the performance of our unconstrained free-placement approach with fixed-site selection, which represents current industry practice. By directly evaluating fixed-site selection, which employs IP-based solvers constrained to existing ground station locations, against the free-placement approach, we can quantify the potential gains achievable when the limitations of pre-defined site locations are removed.

\subsubsection{Synthetic Walker-Star Constellations}
\label{sec:syntheticWalkerStar}

In the unconstrained free-placement problem evaluation, we performed optimizations over a Walker-Star constellation with the parameters listed in \Cref{tab:walkerStar}. This configuration was used to perform parameter sweeps over the total number of satellites in the simulation, ranging from 1 to 4. For our Walker-Star formation, we restricted each orbital plane to contain only one satellite, so the total number of orbital planes $N_\rho$ equals the number of satellites $\lvert\mathcal{S}\rvert$.
\input{walkerConstellation}

\subsubsection{Existing Earth Observation Constellations}
\label{sec:EOconstellations}

For the existing EO satellite constellations considered in this paper, we focused on two major EO operators: Capella Space and ICEYE. The associated orbital characteristics are listed in \Cref{tab:EO_Operators}. We chose these two constellations specifically for the variance in their orbital parameters. Capella Space represents a smaller, mixed constellation of both mid-inclination ($45^\circ$, $53^\circ$) and sun-synchronous ($97^\circ$) orbits, while ICEYE represents a more traditional sun-synchronous-only constellation of a larger size. Examining these two constellations enables us to identify the distinct challenges that arise when designing ground station networks for mixed-orbit versus sun-synchronous trajectories.

\input{satelliteConstellations}

\subsection{Existing Ground Station Networks}

\begin{figure}[htbp!]
    \centering
    \includegraphics[width=0.9\textwidth, trim={30 0 28 15}, clip]{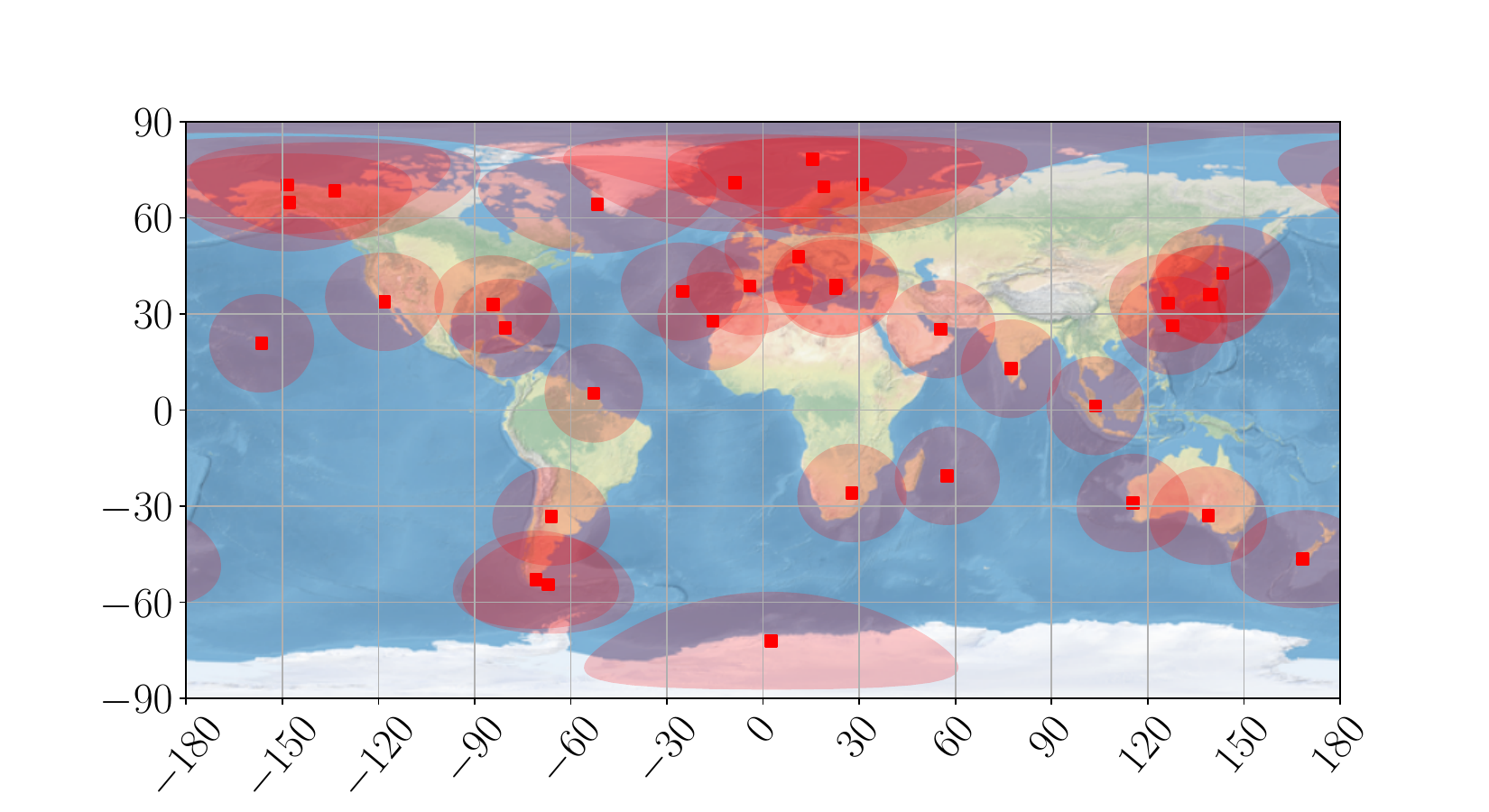} 
    \caption{KSAT Ground Station Network. Locations of operational ground stations from KSAT. Communications cones assume 525km altitude and 10$\degree$ minimum elevation angle.}
    \label{fig:KSAT}
\end{figure}

\begin{figure}[htbp!]
    \centering
    
  \begin{subfigure}[]{0.9\textwidth}
    \centering
    \includegraphics[width=0.9\textwidth, trim={30 0 28 15}, clip]{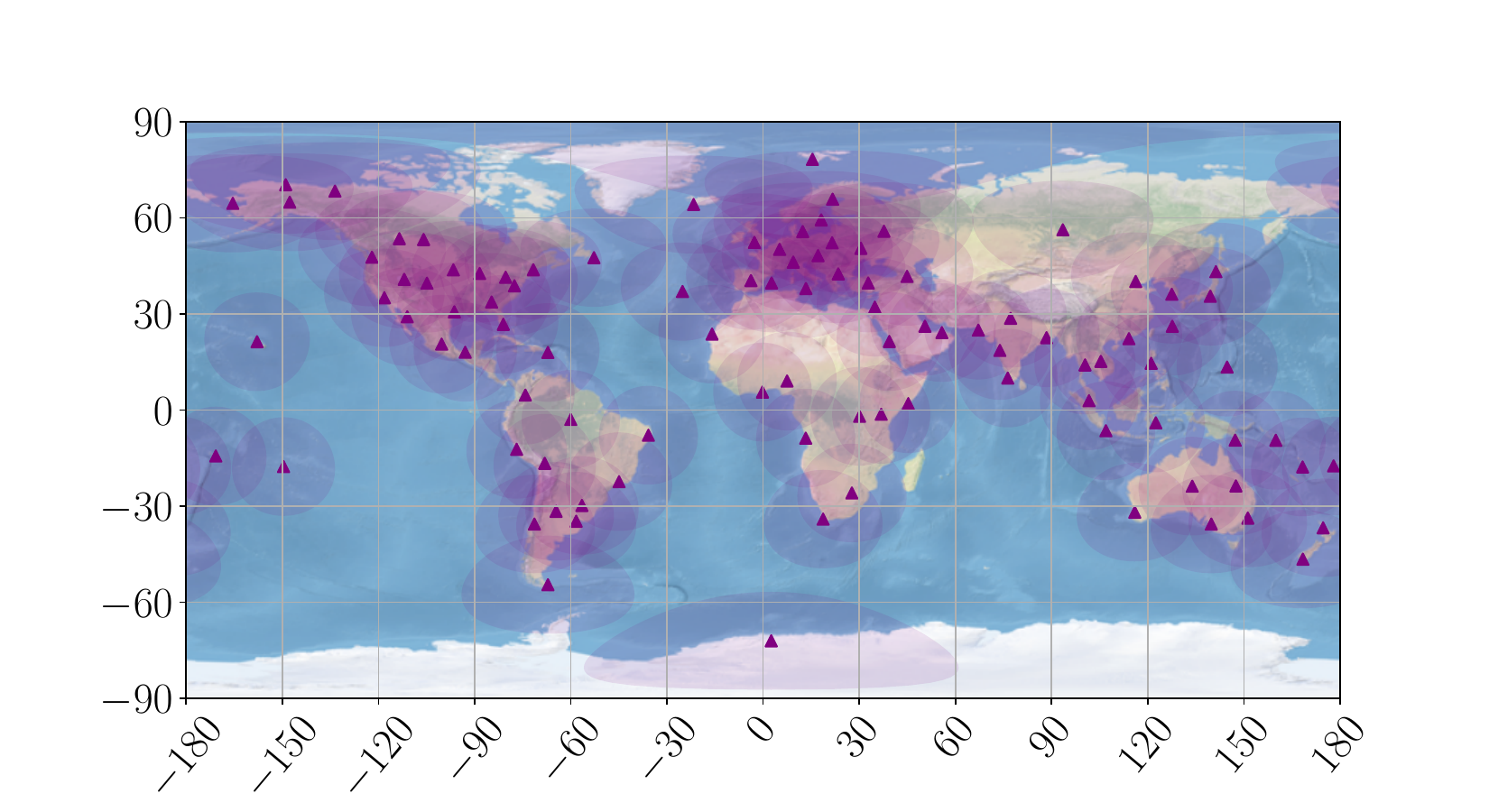}   
    \caption{Global Teleport List. Locations of a geographically diverse subset of 100 teleport stations from the World Teleport Association's global teleport map, selected to maintain computational tractability of the integer programming formulation. Communications cones assume 525km altitude and 10$\degree$ minimum elevation angle.}
    \label{fig:teleportPart}
  \end{subfigure}
  \begin{subfigure}[]{0.9\textwidth}
    \centering
    \includegraphics[width=0.9\textwidth, trim={30 0 28 15}, clip]{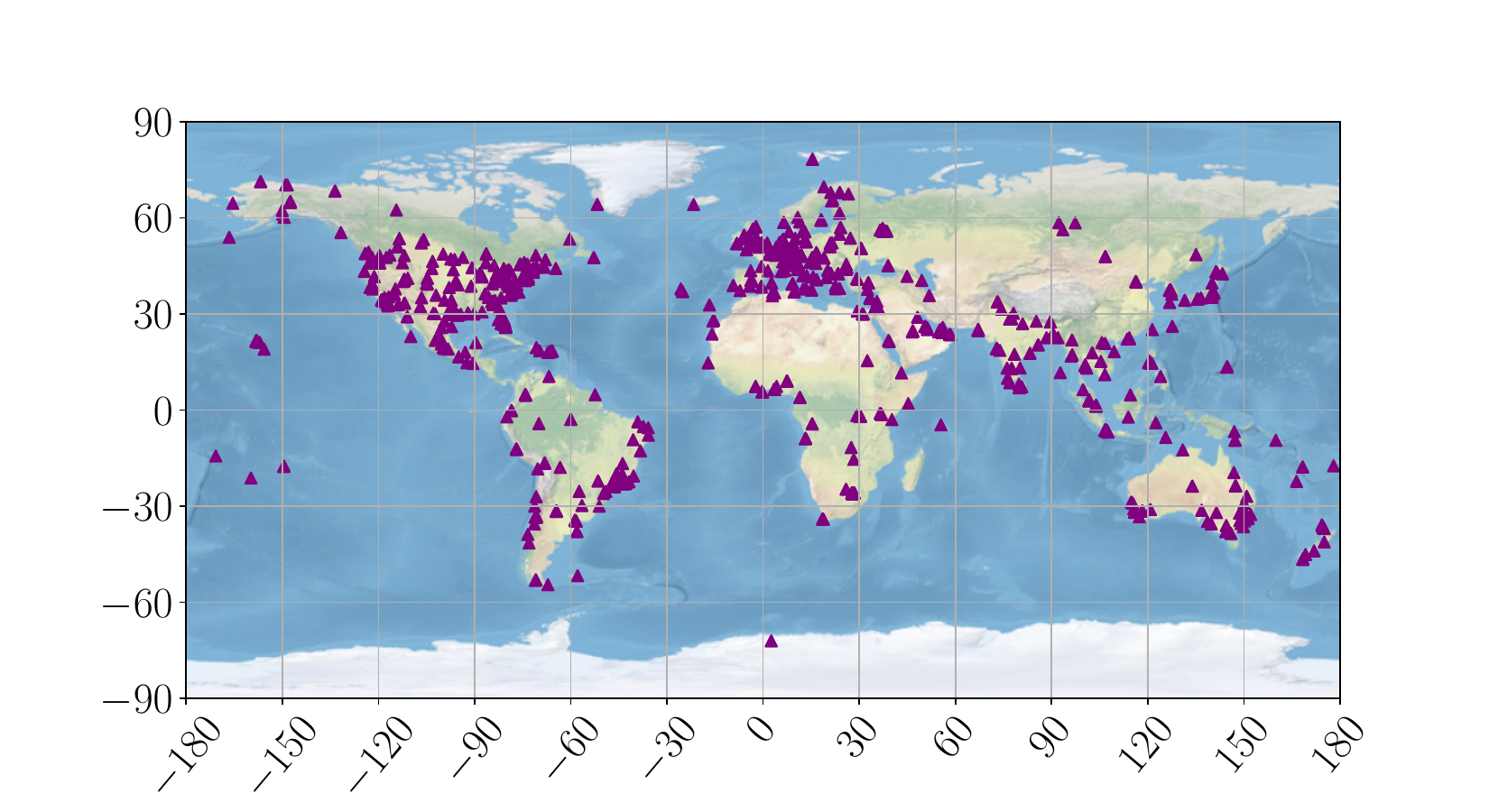}   
    \caption{Full Global Teleport List from the World Teleport Association's global teleport map. Communication cones are not plotted for better visibility of locations.}
    \label{fig:teleportFull}
  \end{subfigure}

  \caption{Global Teleport List Maps.}
    \label{fig:Teleports}
\end{figure}

To compare SCORE against methods that optimize over fixed-site selections, we considered two datasets of ground station locations as potential fixed sites, depicted in \Cref{fig:KSAT,fig:Teleports}. The first dataset is provided by Kongsberg Satellite Services (KSAT), the largest currently existing GSaaS provider. Selecting sites from KSAT’s extensive network represents a particularly challenging GSaaS optimization problem. We also considered a geographically diverse subset of 100 locations from the World Teleport Association's (WTA) global teleport map. The full WTA dataset of 915 stations exceeds the scale at which exact IP solutions can be obtained within a reasonable time, and we therefore selected 100 stations to maximize geographic diversity while maintaining computational tractability. This dataset enables evaluation of network planning in regions with existing satellite communication infrastructure, where operators may wish to expand or optimize ground station deployments even in the absence of a dedicated GSaaS provider. Scaling fixed-site selection to larger candidate sets remains an open challenge, though recent works have begun to address scalable ground station network optimization \cite{kim2026scalable}.

%% file: parameters.tex
\begin{table}[htbp!]
\centering
\addtocounter{table}{-1}
\caption{Simulation Parameters} 
\label{tab:sim_params}
\begin{tabular}{lr}
\toprule
Simulation Parameters & Values\\
\midrule
$t^{start}_{sim}$ & 2025-04-01 17:23:40.69 UTC \\
$t^{end}_{sim}$ & 2025-04-08 17:23:40.69 UTC \\
$t^{start}_{opt}$ & 2025-04-01 17:23:40.69 UTC \\
$t^{end}_{opt}$ & 2026-04-01 17:23:40.69 UTC \\
$C_{dr}$ & \num{1.2} Gbps\\
\bottomrule
\end{tabular}
\end{table}

%% file: walkerConstellation.tex
\begin{table}[htbp!]
\centering
\caption{Walker-Star Constellation Parameters} 
\label{tab:walkerStar}
\begin{tabular}{lr}
\toprule
Walker-Star Constellation Parameters & Values\\
\midrule
Altitude & 781 km\\
Eccentricity& 0.001 \\
Inclination& 86.4\textdegree \\
Number of Planes Varied in Parameter Sweep ($N_\rho$) & 1--4\\
Satellites per Plane ($s_{\text{per plane}}$) & 1\\
\bottomrule
\end{tabular}
\end{table}

%% file: satelliteConstellations.tex
\begin{table}[htbp!]
\centering
\caption{Earth Observation Satellite Constellations} 
\label{tab:EO_Operators}
\begin{tabular}{lrrr}
\toprule
EO Satellite Operators & Constellation Size & Orbital Altitudes & Inclination Angles\\
\midrule
Capella Space & 5 & 525-575 km & $45^{\circ}$, $53^{\circ}$, $97^{\circ}$\\
ICEYE & 34 & 560-580 km & $97^{\circ}$\\
\bottomrule
\end{tabular}
\end{table}





%% file: 06_Results.tex
\section{Results}
We organize the results into two main evaluations. First, we compare SCORE with differential evolution to assess computational efficiency, convergence, and the potential for multiple near-optimal solutions using a synthetic Walker-Star constellation. This analysis quantifies SCORE’s convergence speed improvements relative to DE, showing that SCORE achieves equivalent or superior objective values while reducing the number of function evaluations to convergence by up to 5$\times$ fewer function evaluations than DE for larger ground station selection configurations.

Next, we evaluate SCORE against fixed-site selection methods based on integer programming (IP) to analyze performance tradeoffs when network locations are constrained to existing infrastructure. Using the KSAT and WTA teleport networks as baselines, we quantify how limited site availability restricts achievable downlink performance compared to SCORE’s unconstrained placement. Focusing on site placement for the CAPELLA and ICEYE constellations, we found that SCORE consistently outperformed IP-based fixed-site methods, achieving improvements in data throughput of over 9\% relative to fixed-site solutions.  Collectively, these evaluations highlight SCORE’s scalability, efficiency, and applicability across a variety of ground station network design scenarios.

\subsection{Free Placement Problem}
\label{sec:SCOREVSDE}
We begin by comparing methods for the free-placement problem, benchmarking SCORE against differential evolution using the synthetic Walker-Star constellation described in \Cref{sec:syntheticWalkerStar}. The mission objective $f$ was to maximize total data downlink over the simulation period, $T_{sim}$ as outlined in \Cref{eqn:obj_max_data}. Final objective function values were scaled by $\frac{T_{opt}}{T_{sim}}$ to approximate the data downlink over entire optimization period. We varied the number of satellites and ground stations independently over $\{1,2,3,4\}$, generating 16 unique scenarios to evaluate algorithmic performance and computational efficiency. Our evaluations focused on the number of function evaluations to reach convergence and solution quality, with each scenario repeated over 10 random seeds to ensure robustness.  Contact exclusion constraints were handled in two steps to prevent overlapping satellite contacts across ground stations. After optimization, solutions were post-processed with an integer programming solver to enforce strict non-overlapping assignments, counting any overlaps only once in the final objective.

\begin{figure}[htbp!]
  \centering

  \begin{subfigure}[]{0.75\textwidth}
    \centering
    \includegraphics[page=1,width=\textwidth]{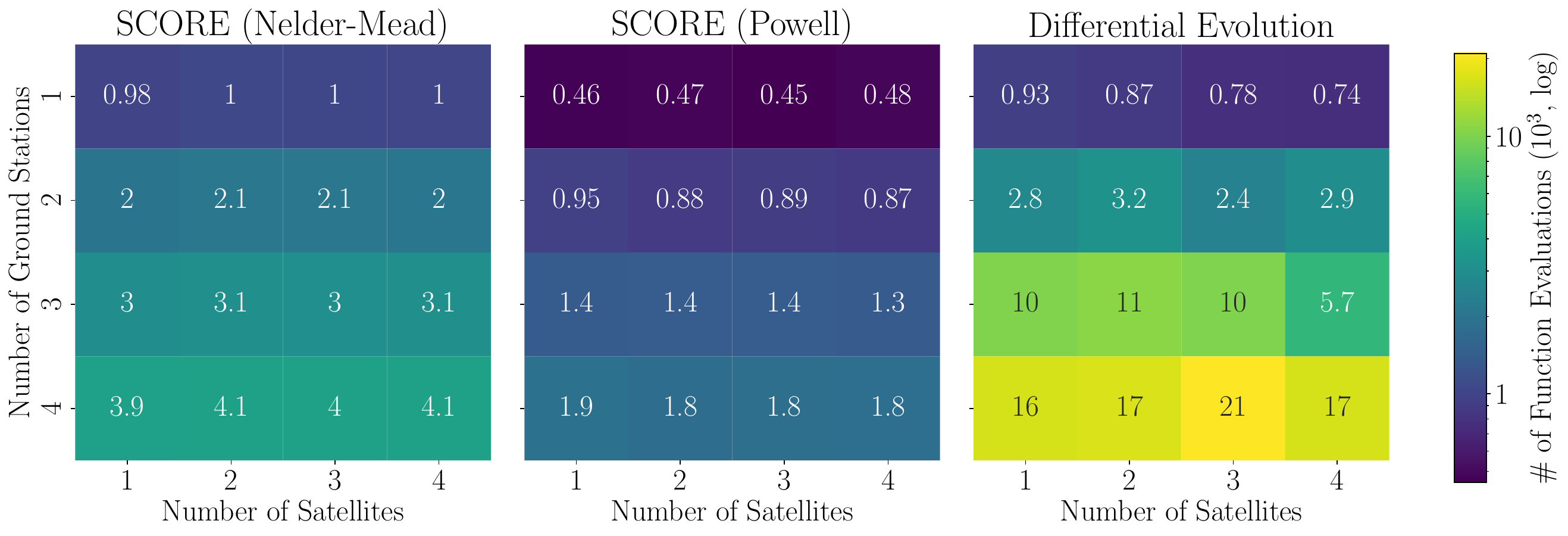}
    \caption{Total number of function evaluations to reach convergence for SCORE (Nelder-Mead and Powell) and DE free placement algorithms}
    \label{fig:heatmap_comp}
  \end{subfigure}
  \begin{subfigure}[]{0.75\textwidth}
    \centering
    \includegraphics[page=1,width=\textwidth]{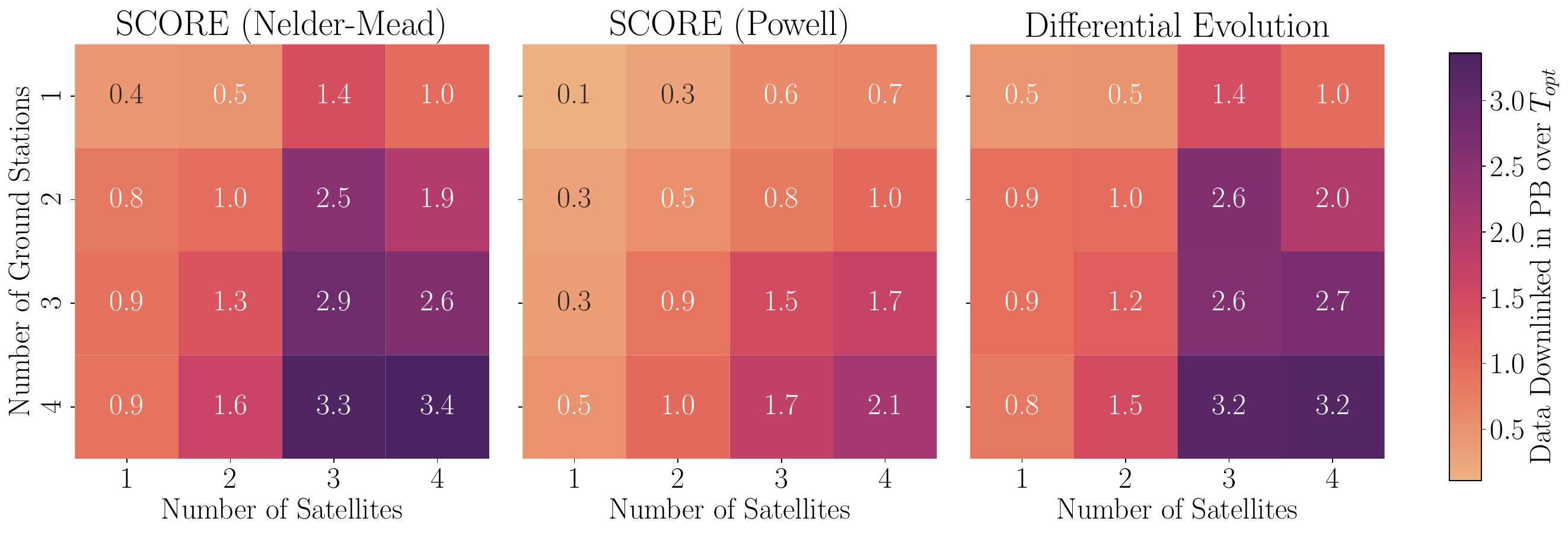}
    \caption{Final data downlinked in PB for SCORE (Nelder-Mead and Powell) and DE free placement methods}
    \label{fig:heatmap_obj}
  \end{subfigure}

  \caption{Computation versus performance gains for SCORE versus leading free placement optimization algorithms}
  \label{fig:heatmap_comparison}
\end{figure}
\subsubsection{Computation and Performance Gains}
First, we evaluated SCORE and DE’s tradeoffs between computation efficiency and mission objective performance. \Cref{fig:heatmap_comparison} displays two critical metrics: (a) the number of function evaluations required for each algorithm to reach solution convergence and (b) the final objective values of total data downlinked in petabytes over $T_{opt}$. 
We evaluate SCORE using two derivative-free local search methods, Nelder-Mead and Powell, demonstrating optimizer flexibility while maintaining lower per-iteration cost than global meta-heuristic methods such as DE. In the heatmaps of \Cref{fig:heatmap_comp}, SCORE’s number of function evaluations increases linearly with the number of ground stations $n$ due to its sequential selection process. However, DE’s number of function evaluations grows much more rapidly, reaching over 17,000 function evaluations in the largest scenario compared to SCORE’s 4,100. For both algorithms, varying the constellation size $\lvert\mathcal{S}\rvert$ had minimal effect on number of function evaluations needed until convergence. 

\Cref{fig:heatmap_obj} shows the final mission objective values of total data downlinked in petabytes for both methods. Crucially, SCORE's efficacy is sensitive to the choice of underlying optimizer; while the Powell method offers the lowest computational overhead (averaging roughly 50\% fewer evaluations than Nelder-Mead), it results in significantly degraded objective values, often failing to surpass DE performance on even the simplest of configurations. Conversely, using a different optimizer like Nelder-Mead allows SCORE to maintain its predictable linear growth in complexity while consistently attaining near-optimal downlink performance. In larger networks, SCORE with Nelder-Mead surpasses DE by as much as 13\% in total data downlinked (1 satellite, 4 ground stations), despite DE requiring up to five times more function evaluations (e.g., 21,000 vs 4,000 evaluations at 4 ground stations and 3 satellites). These results highlight SCORE’s superior efficiency, scalability, and effectiveness in exploring the solution space for ground station network optimization, provided a high-performing local search method is employed.

\begin{figure}[htbp!]
    \centering
    \includegraphics[width=\linewidth]{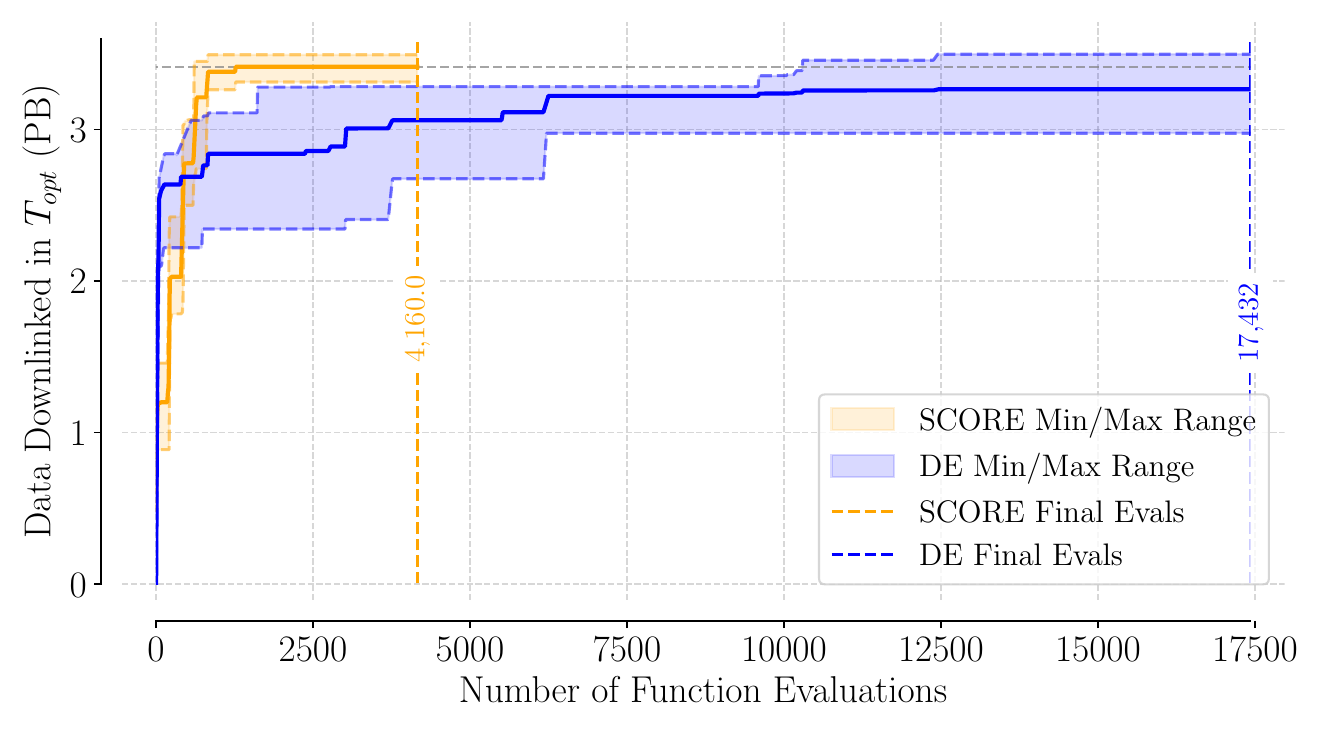}
    \caption{Data downlink comparison for SCORE and DE. SCORE (orange) converges towards a solution in around 4000 function evaluations; DE (blue) converges more slowly with higher variability.}
    \label{fig:Convergence_Score_DE}
\end{figure}

\subsubsection{Convergence Behavior}
To better understand differences in convergence behavior between SCORE and DE, \Cref{fig:Convergence_Score_DE} shows performance over time for the largest scenario from \Cref{fig:heatmap_comparison}: a four-satellite Walker-Star constellation requiring placement of four ground stations. We plot the total data downlinked against the number of function evaluations completed by each method. SCORE quickly reaches near-maximum downlink, achieving a mean performance of 3.4 PB in an average of 4,160 function evaluations (orange curve, vertical dashed line). DE’s increase is slower and more variable, reaching only 3.2 PB after an average of 17,432 function evaluations (blue dashed line).

SCORE’s fast convergence can be attributed to its strategic sequential coordinate selection phase and Nelder-Mead’s efficient simplex search, which require few function evaluations to iteratively improve solutions. After fast selection of an initial set of coordinates, the subsequent cyclic refinement facilitates incremental improvements until convergence, finding solutions with up to 5$\times$ fewer function evaluations than DE, while achieving comparable solution quality. The shaded regions surrounding each algorithm’s average convergence curve reflect variability across random seeds. SCORE’s variability remains limited, while DE’s stochastic, population-based search manifests in broader variability and less predictable convergence rates, due to its reliance on initial populations and random genetic operations. For further ablation studies, we direct the reader to Appendix~\ref{app:DESweep}, where we explore a wider variety of DE parameters, and Appendix~\ref{app:score_ablation}, where we validate the cyclic refinement step and its robustness to selection ordering, both for the largest scenario of 4 ground stations and 4 satellites.

\subsubsection{Multiple Near-Optimal Solutions}

\begin{figure}[htbp!]
  \centering

  \begin{subfigure}[t]{0.6\textwidth}
    \centering
    \includegraphics[width=\textwidth, trim={30 20 28 15}, clip]{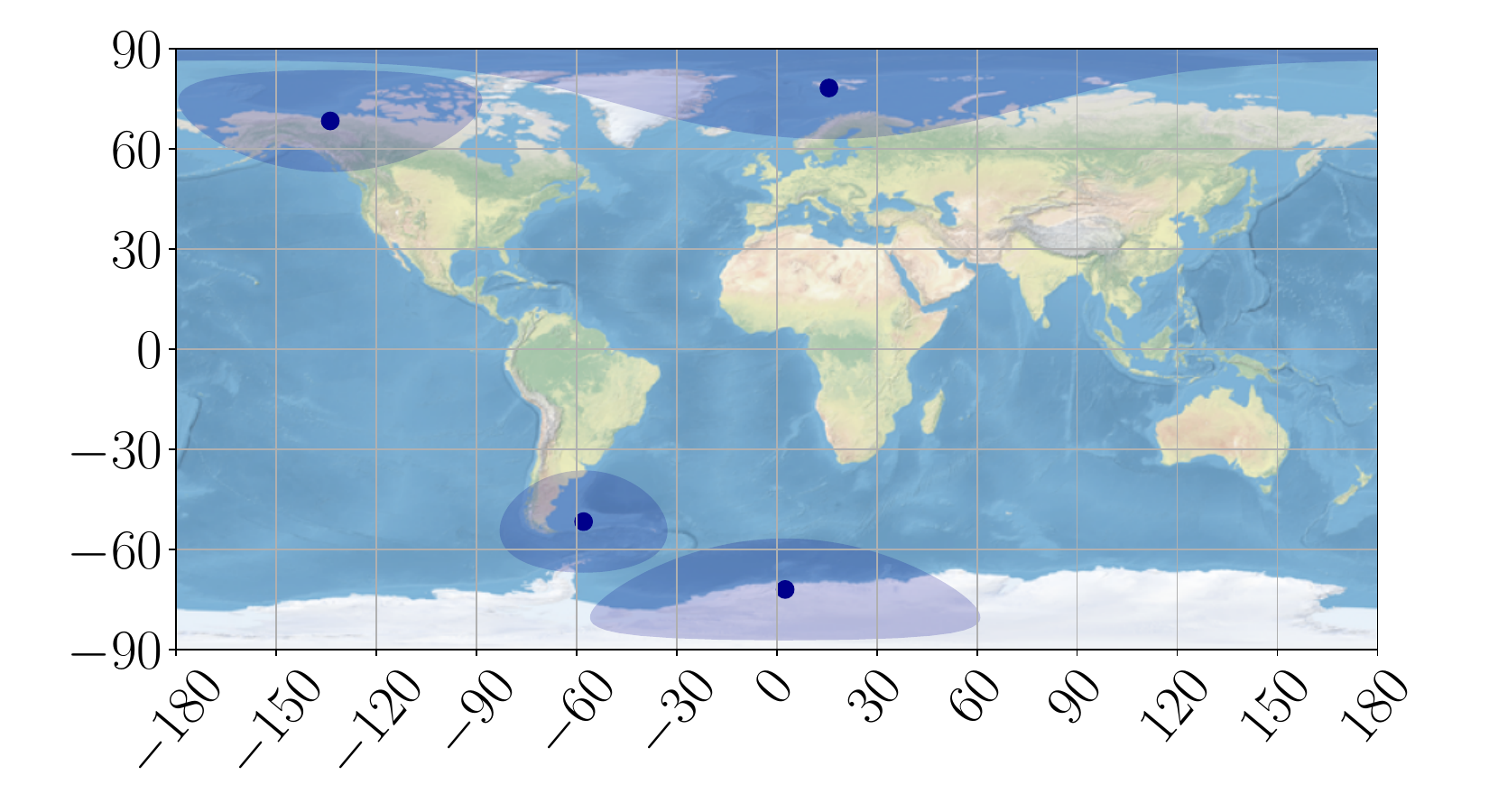}
    \caption{}
    \label{fig:multiplesol1}
  \end{subfigure}

  \vspace{1em}

  \begin{subfigure}[t]{0.49\textwidth}
    \centering
    \includegraphics[width=\textwidth, trim={30 20 28 15}, clip]{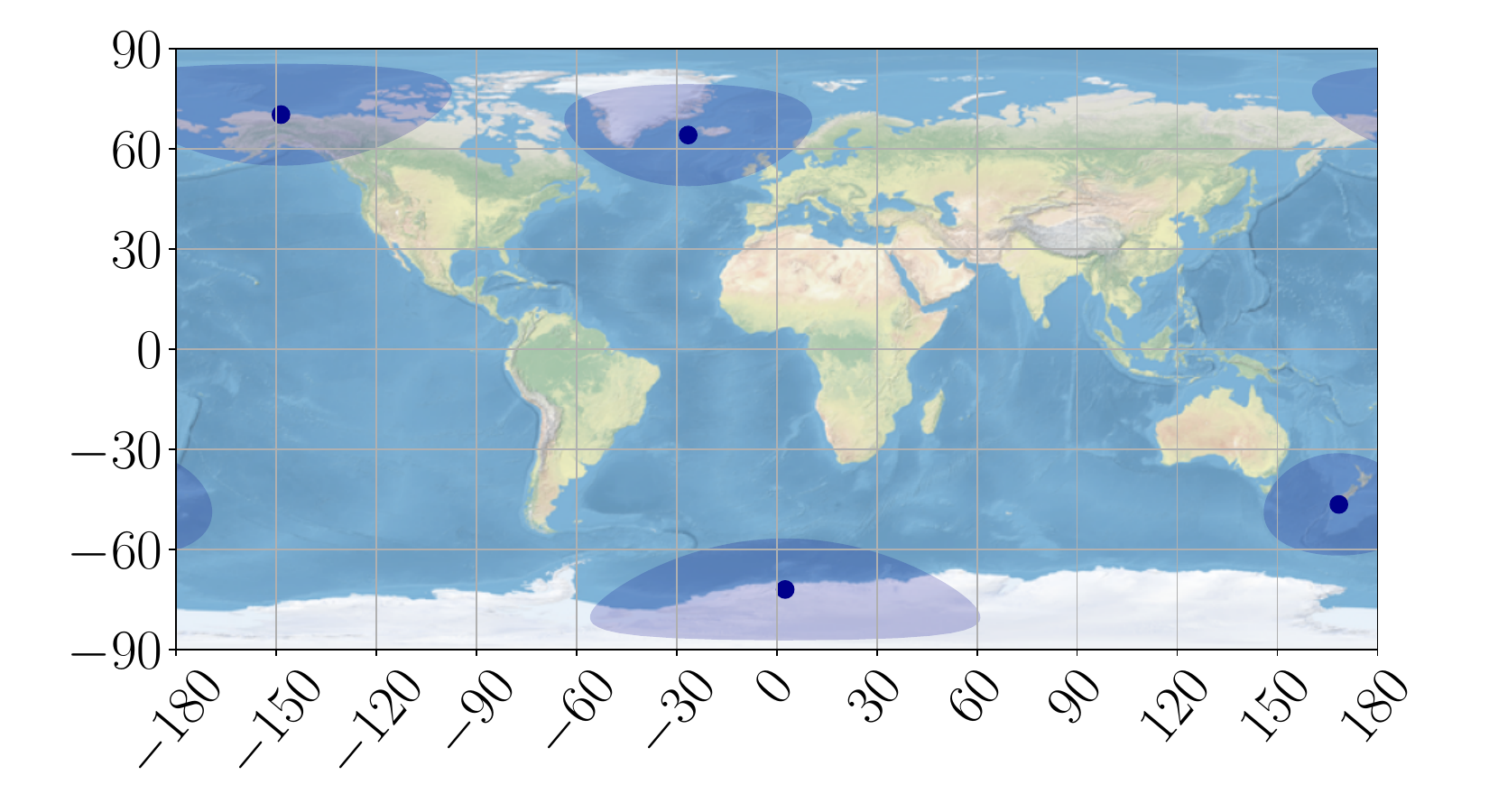}
    \caption{}
    \label{fig:multiplesol2}
  \end{subfigure}
  \hfill
  \begin{subfigure}[t]{0.49\textwidth}
    \centering
    \includegraphics[width=\textwidth, trim={30 20 28 15}, clip]{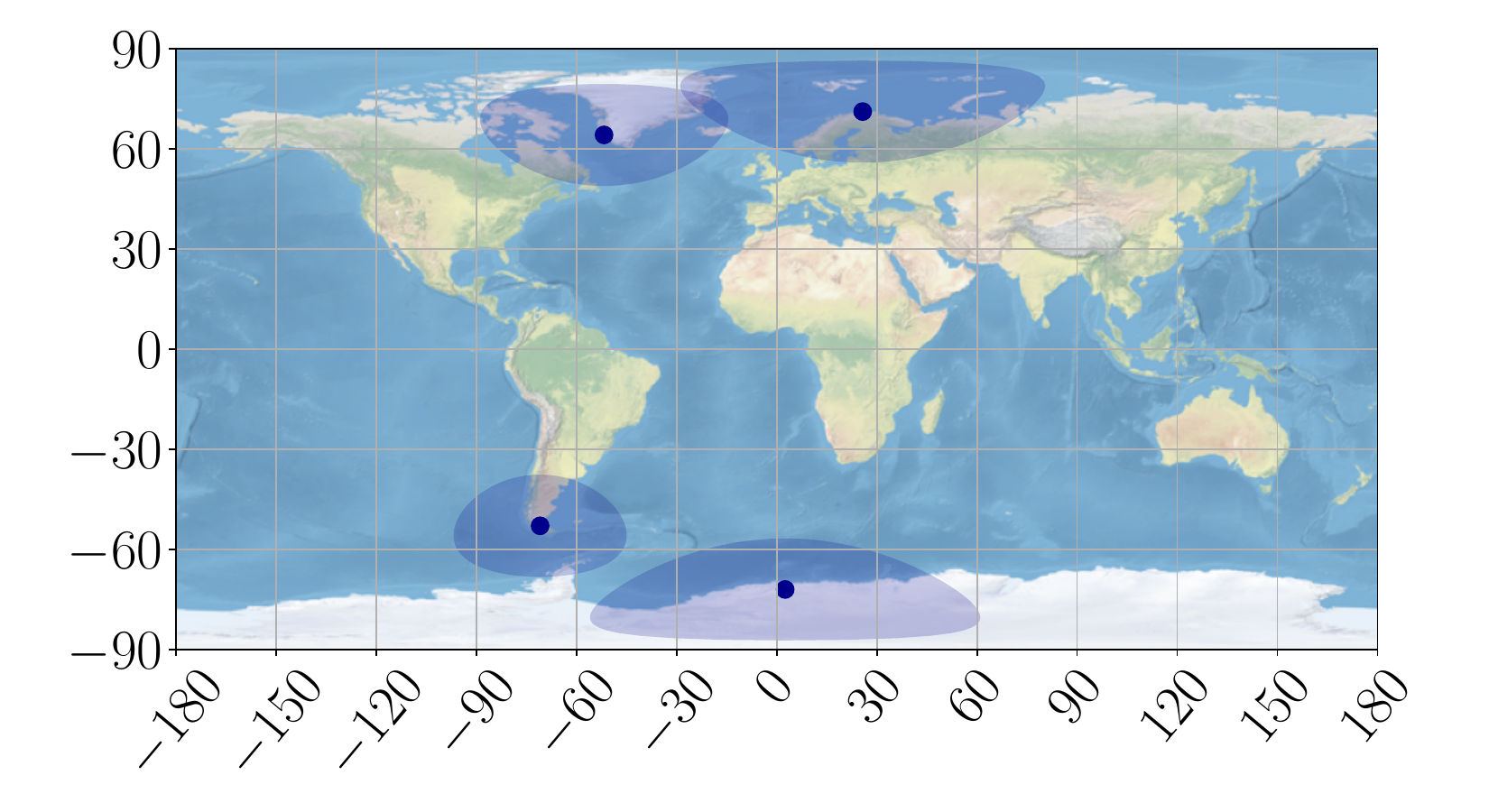}
    \caption{}
    \label{fig:multiplesol3}
  \end{subfigure}

  \caption{Multiple high-performing ground station layouts yield similar objectives, showing placement flexibility. Coordinates and objective values are listed in \Cref{tab:multi_solution_coordinates}.}
  \label{fig:multipleSolutions}
\end{figure}

In experiments varying random seeds, we found multiple distinct ground station layouts for the four-satellite Walker-Star constellation with four stations. Several example arrangements are displayed in \Cref{fig:multipleSolutions}. Despite differences in spatial arrangement, all configurations achieved similar objective values, indicating multiple local optima. This suggests that overall layout patterns, rather than exact station positions, are the key factors influencing performance. The full list of coordinates and their data downlink values over $T_{opt}$ is provided in \Cref{tab:multi_solution_coordinates}.

\input{multiSolutions}

\subsection{Free Placement versus Fixed-Site Selection Methods}
\label{sec:freevsFixed}
We next evaluate SCORE's performance against fixed-site selection optimization using integer programming solvers. To introduce greater realism, we compare SCORE to IP optimizations over existing ground station networks, specifically those provided by KSAT and WTA's global teleport map. We evaluate three variants of SCORE against our fixed-site methods. First, unconstrained SCORE performs free placement with no infrastructure penalties, serving as a strong empirical performance benchmark on achievable downlink performance. Second, SCORE (Lat-Const) restricts placement to latitudes within $[-73^\circ, 78^\circ]$, corresponding to the range of existing KSAT infrastructure, representing a middle ground between fully unconstrained and infrastructure-constrained optimization. Third, SCORE (Infra-Const) incorporates the infrastructure proximity penalty from \Cref{eqn:penalty_infra}, encouraging placement near existing population centers with access to fiber and power infrastructure. Comparing these three variants isolates the contribution of purely geometric optimization from the practical costs of site feasibility, quantifying the performance tradeoff operators face when moving from idealized to operationally realistic deployments. We further focus our evaluation on maximizing data downlink for established Earth observation satellite constellations, namely Capella Space and ICEYE, with their characteristics detailed in \Cref{sec:EOconstellations}. When optimizing over a predefined set of candidate locations $\mathcal{P}$, the IP solver returns a globally optimal solution, assuming a feasible solution exists.


\subsubsection{Performance between SCORE, Fixed-Site Selection from Teleports and KSAT}
\begin{figure}[htbp!]
  \centering

  \begin{subfigure}[t]{0.49\textwidth}
    \centering
    \includegraphics[width=\textwidth]{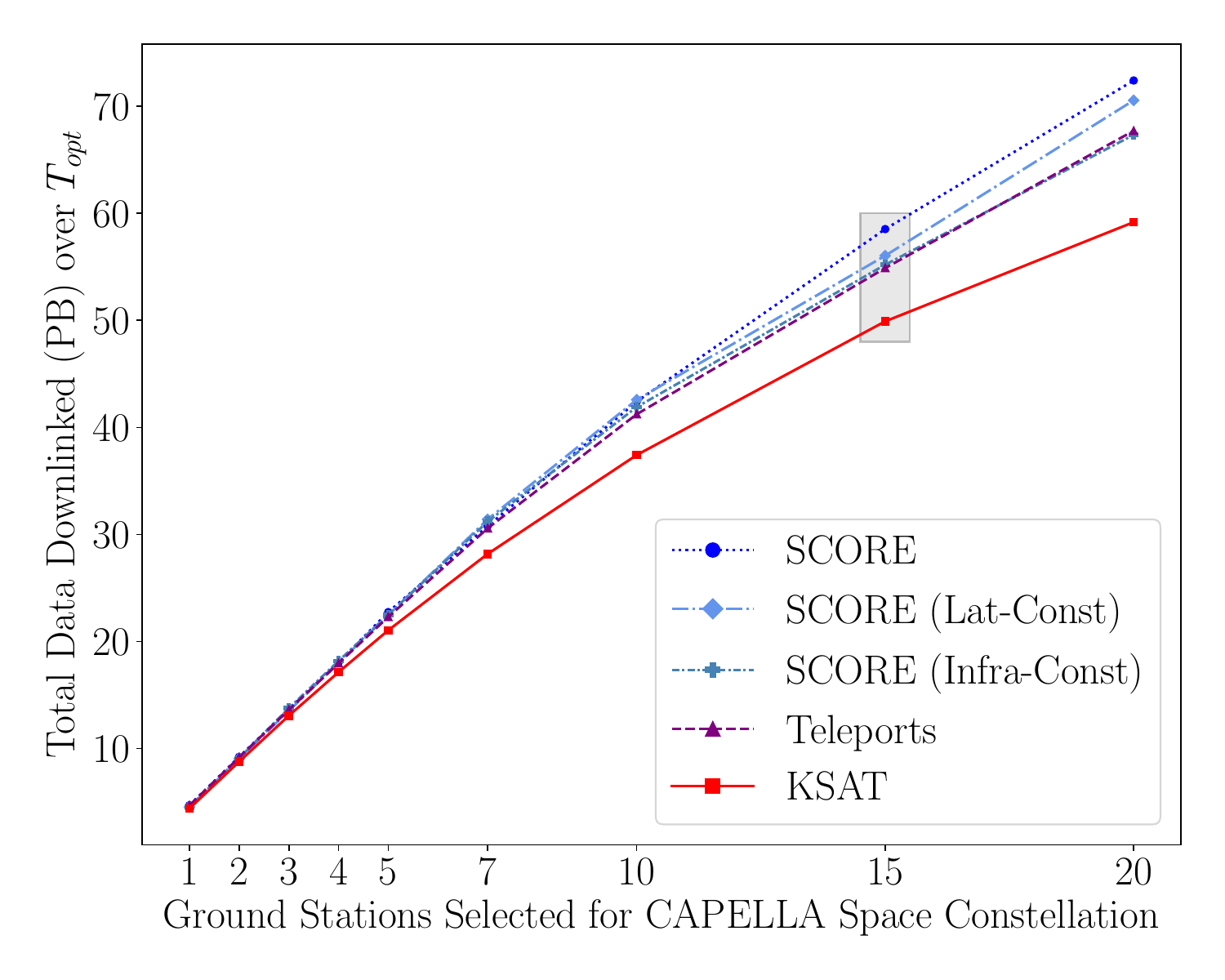}
    \caption{Full Capella Space Constellation}
    \label{fig:capella_full}
  \end{subfigure}
  \hfill
  \begin{subfigure}[t]{0.49\textwidth}
    \centering
    \includegraphics[width=\textwidth]{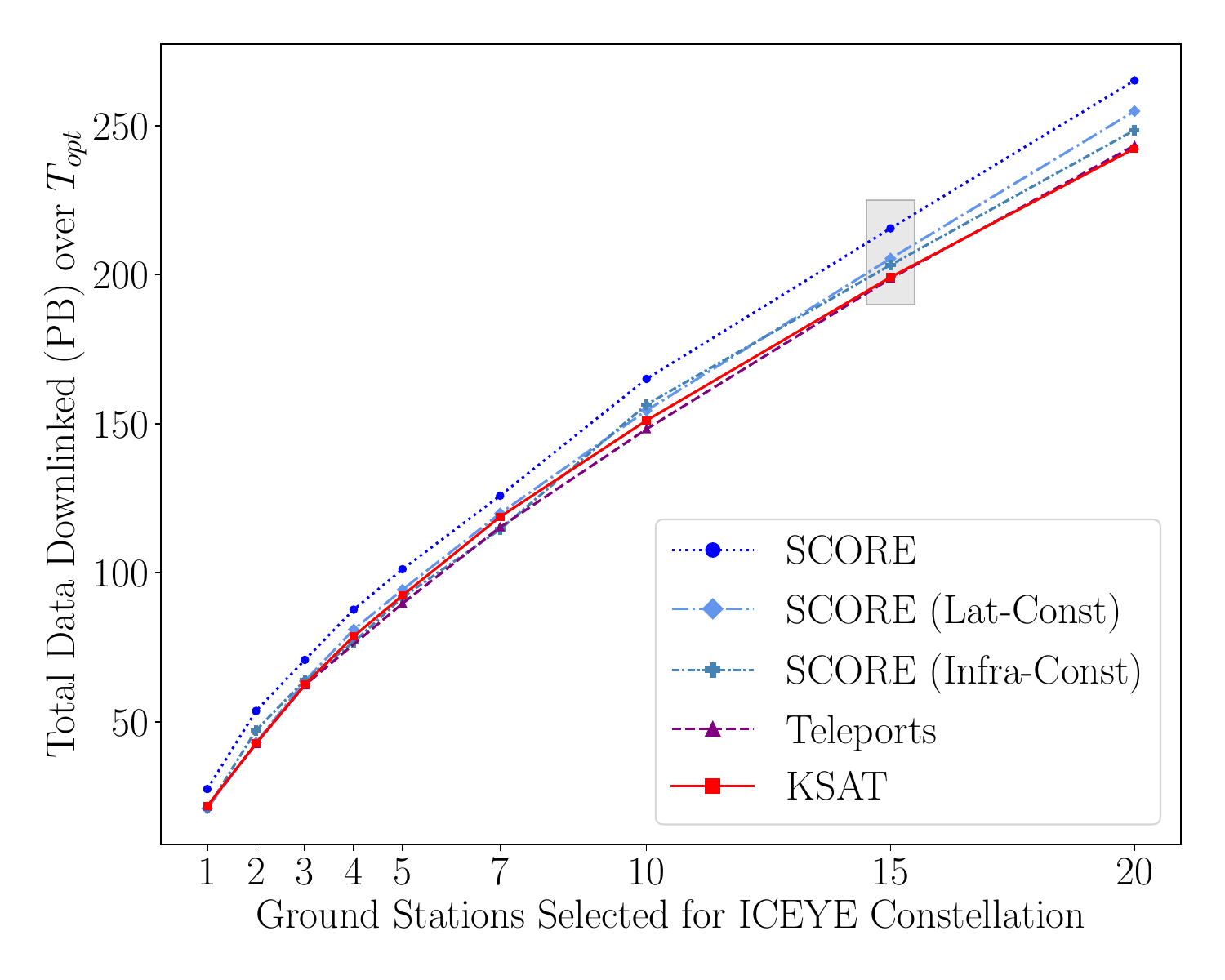}
    \caption{Full ICEYE Constellation}
    \label{fig:iceye_full}
  \end{subfigure}

  \vspace{1em}

  \begin{subfigure}[t]{0.49\textwidth}
    \centering
    \includegraphics[width=\textwidth]{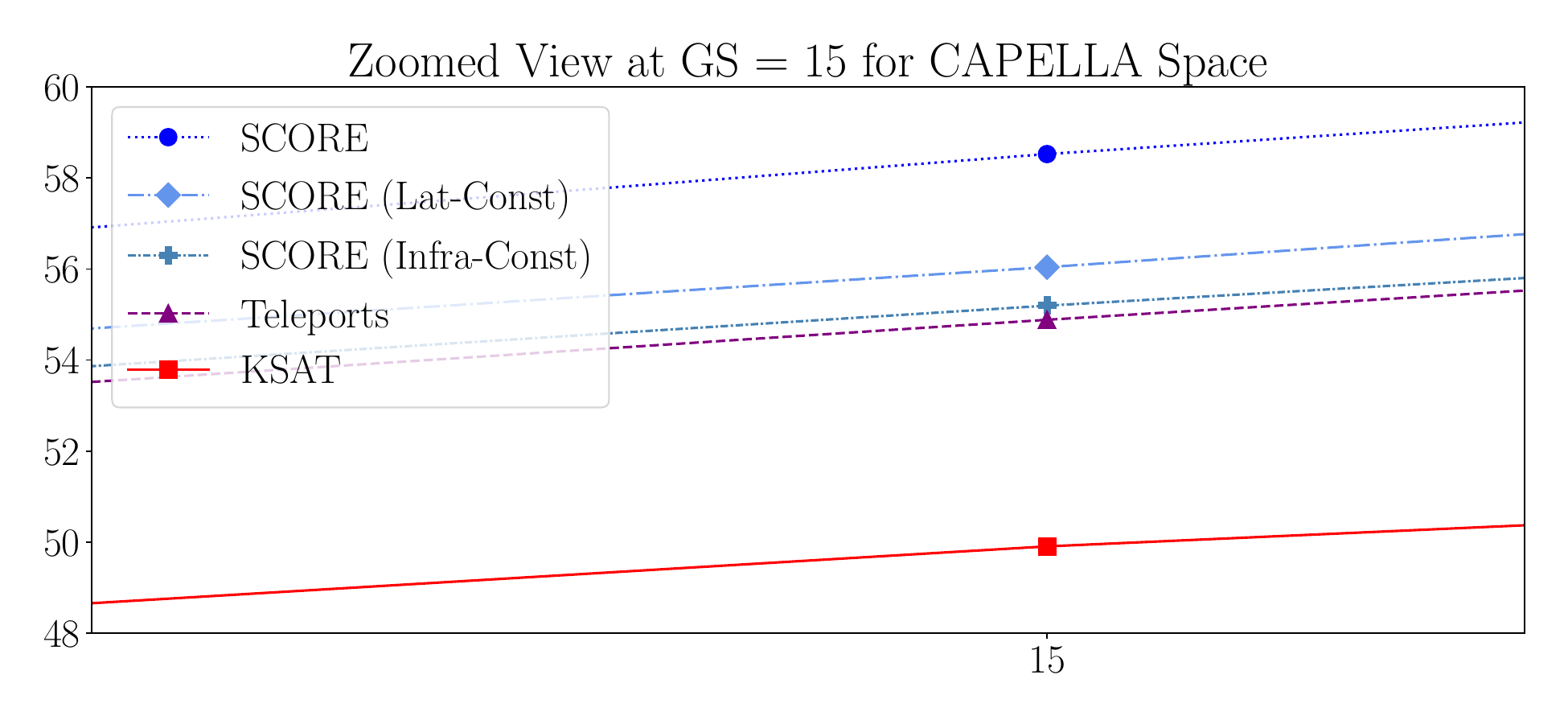}
    \caption{Zoomed View of Capella Constellation}
    \label{fig:capella_zoom}
  \end{subfigure}
  \hfill
  \begin{subfigure}[t]{0.49\textwidth}
    \centering
    \includegraphics[width=\textwidth]{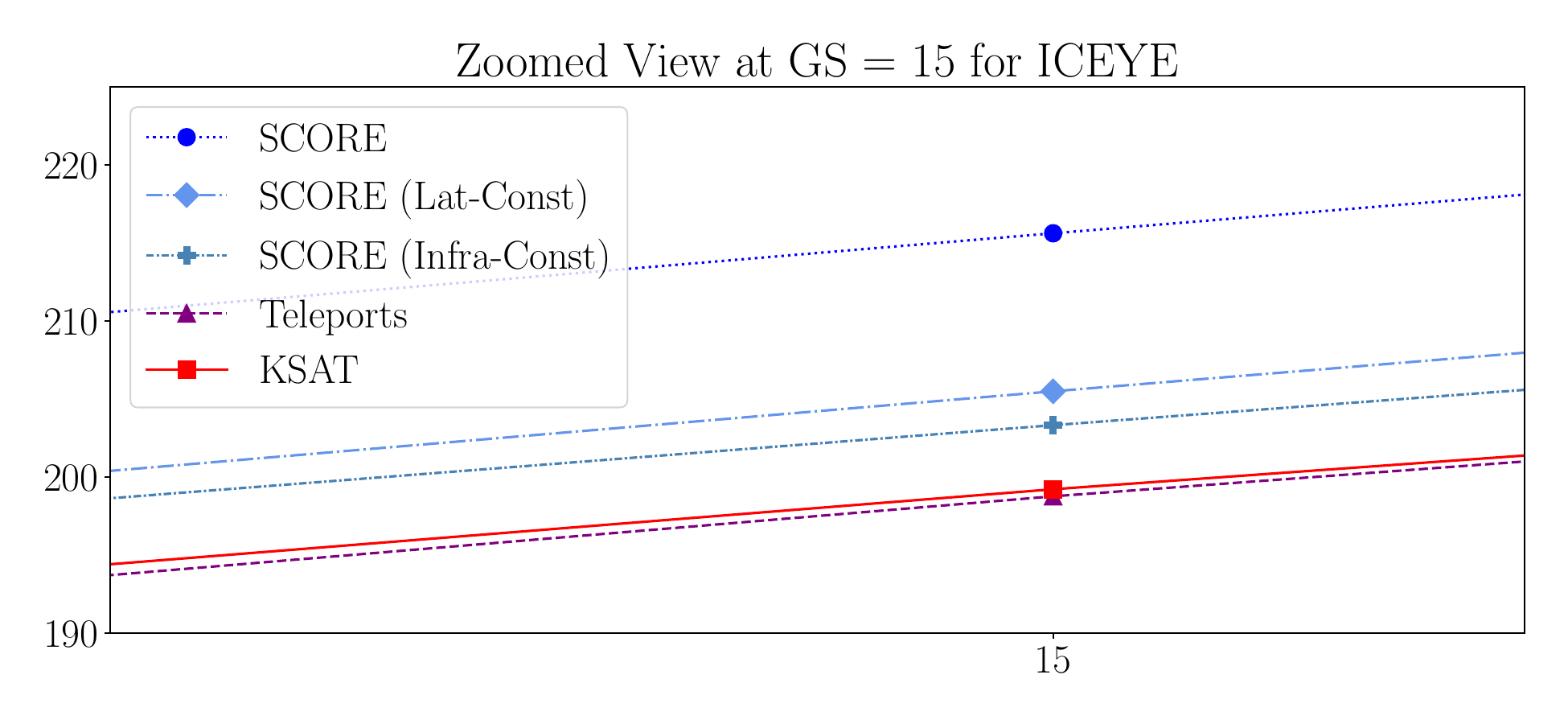}
    \caption{Zoomed View of ICEYE Constellation}
    \label{fig:iceye_zoom}
  \end{subfigure}

  \caption{Comparison of full constellation views (top) and their corresponding zoomed-in regional views (bottom) for Capella and ICEYE constellations, illustrating free-placement vs. fixed-site ground station selection.}
  \label{fig:capella_iceye_free_fixed_comparison}
\end{figure}

The downlink performance of SCORE and IP-based optimization over the mission window $T_{opt}$ is shown in \Cref{fig:capella_iceye_free_fixed_comparison} for networks with 1 to 20 stations. For the Capella Space constellation (\Cref{fig:capella_full}), unconstrained SCORE consistently outperforms IP-based methods, although IP achieves similar results for smaller networks. The strong performance of the fixed-site IP solutions can be attributed to Capella’s mid-latitude orbit inclinations, which align well with the predominantly mid-latitude distribution of ground stations and teleports shown in \Cref{fig:KSAT,fig:Teleports}. In this case, unconstrained SCORE yields only marginal gains for small networks, 0.4\% over teleports and 4.6\% over KSAT for three stations, but its advantage grows with network size, reaching improvements of 7\% over teleports and 22\% over KSAT at 20 stations. For the ICEYE constellation (\Cref{fig:iceye_full}), unconstrained SCORE significantly outperforms both IP baselines, leading by 8–15\% across all network sizes. This reflects the limited high-latitude ground stations in KSAT and teleport networks, which hinders support for ICEYE’s high-inclination orbits. 
We note that these unconstrained results represent a strong empirical performance benchmark, as this version of SCORE is free to select locations regardless of infrastructure feasibility or operational accessibility.

The constrained SCORE variants reveal the practical cost of enforcing infrastructure feasibility. For Capella, both SCORE (Lat-Const) and SCORE (Infra-Const) perform slightly below unconstrained SCORE but remain close to the Teleports IP baseline, as Capella's mid-latitude orbits are well-served by infrastructure-accessible locations. Notably, SCORE (Infra-Const) tracks closely with the Teleports line, suggesting that for mid-latitude constellations, restricting placement to infrastructure-accessible regions yields performance comparable to existing teleport networks. For ICEYE, the latitude constraint incurs a more meaningful performance penalty. At small network sizes, constrained SCORE variants perform comparably to fixed-site methods, as the most geometrically valuable polar locations are excluded from the search space. For larger networks, however, SCORE (Lat-Const) and SCORE (Infra-Const) consistently outperform both Teleports and KSAT, achieving gains of 2--5\% over Teleports and 2--5\% over KSAT at 20 stations. This compares to unconstrained SCORE's 8--15\% improvement, indicating that infrastructure constraints reduce but do not fully eliminate the performance advantage of free placement. These results highlight a key tradeoff for operators: infrastructure constraints reduce the performance ceiling of free placement, particularly for high-inclination constellations, but provide more physically attainable site placements that still can outperform fixed-site selections at larger network sizes.

\subsubsection{Coordinate Location Comparisons}

\begin{figure}[htbp!]
    \centering
    \vspace{-1em}
    \includegraphics[width=0.73\textwidth, trim={30 0 28 15}, clip]{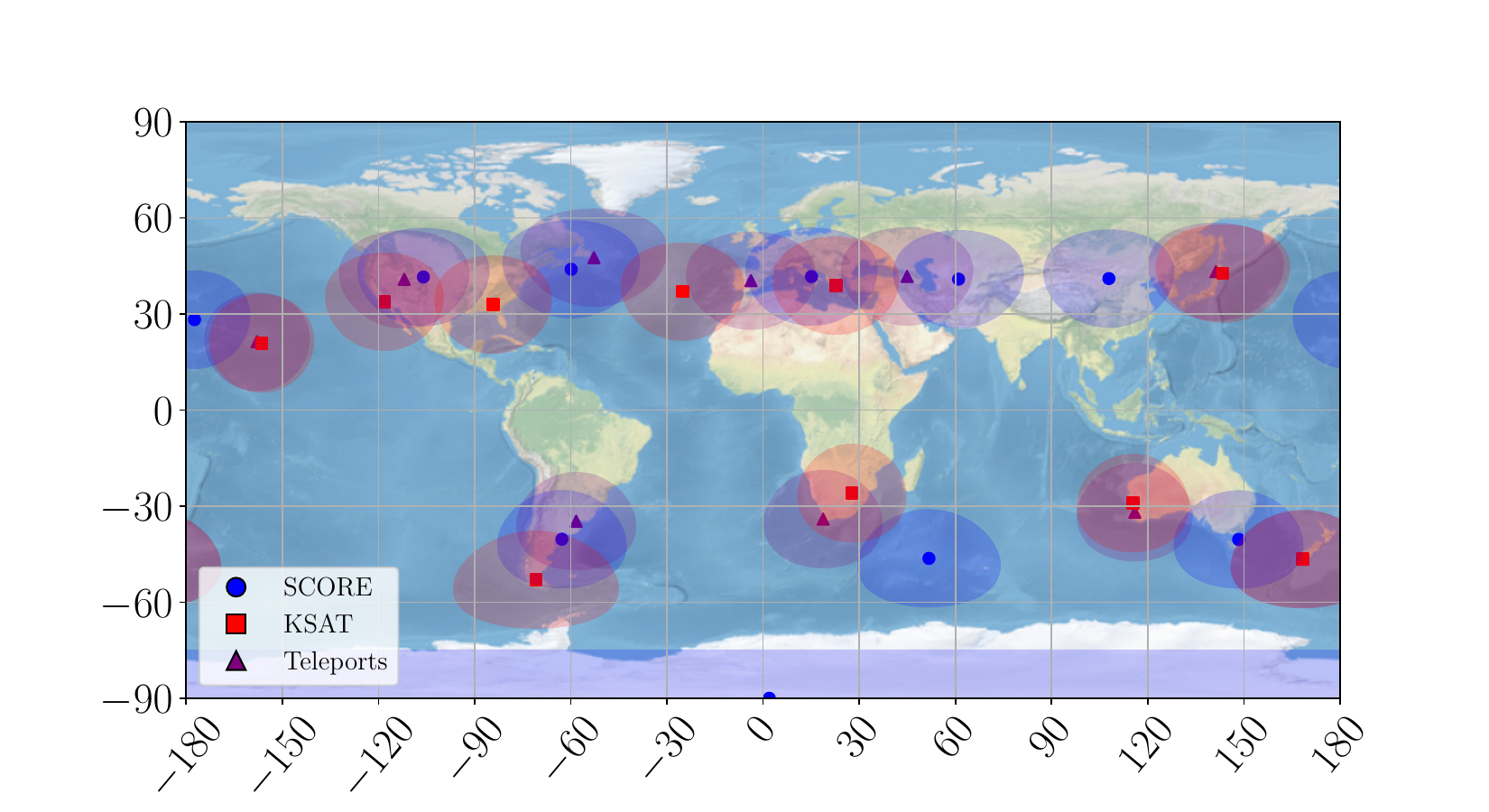} 
    \caption{Optimization over Capella Space Constellation showing final ground station locations for a network of 10 stations. Solutions include KSAT stations (red), Teleport sites (purple), and SCORE free-placement (blue).}
    \label{fig:Capella_locations_10}
\end{figure}

\begin{figure}[htbp!]
    \centering
    \vspace{-1em}
    \includegraphics[width=0.73\textwidth, trim={30 0 28 15}, clip]{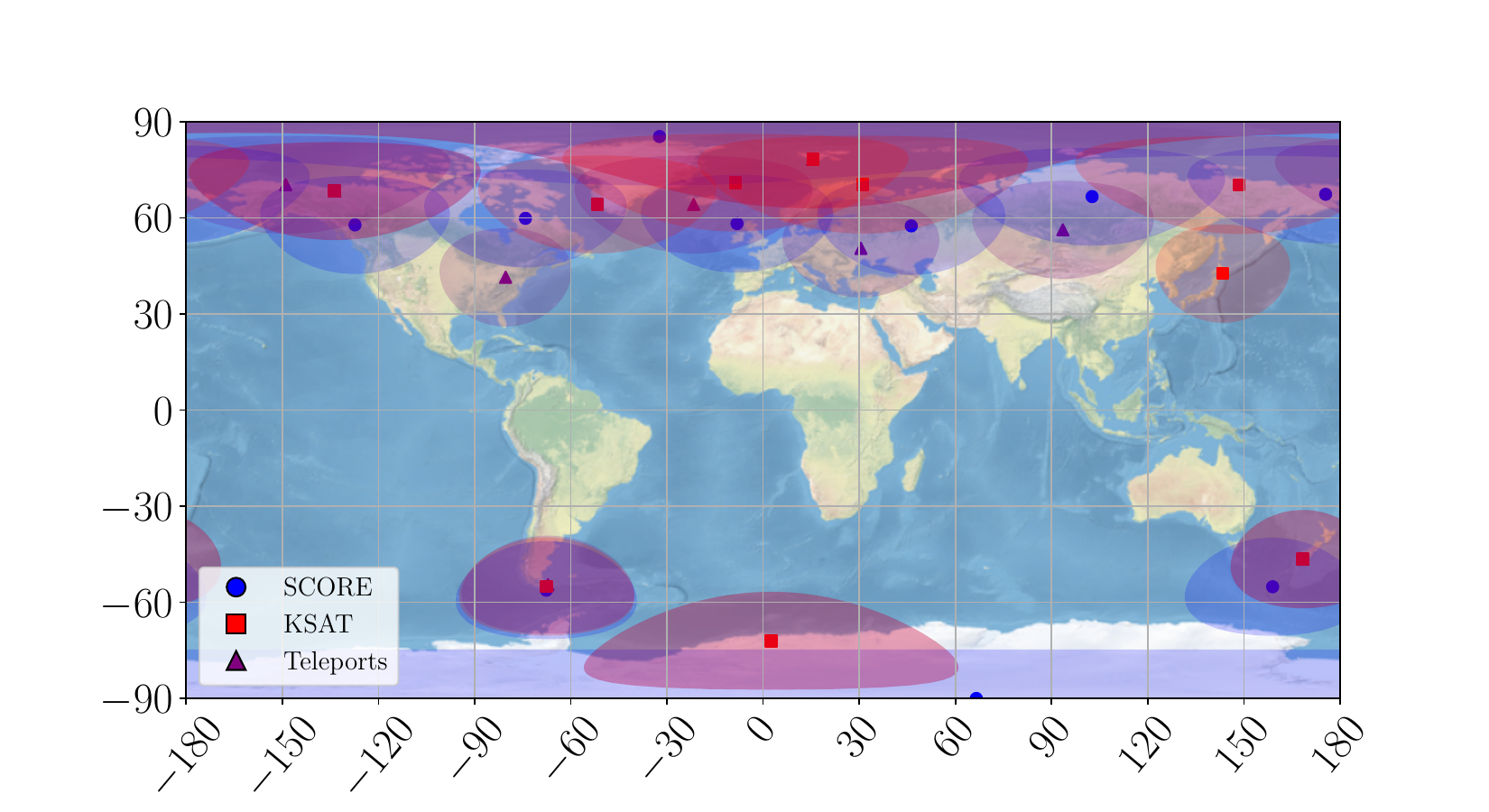}  
    \caption{Optimization over ICEYE Constellation showing final ground station locations for a network of 10 stations. Solutions include KSAT stations (red), Teleport sites (purple), and SCORE free-placement (blue).}
    \label{fig:ICEYE_locations_10}
\end{figure}

We now perform a detailed evaluation of optimized ground station coordinates for a 10-station network of both the Capella Space (\Cref{fig:Capella_locations_10}) and ICEYE (\Cref{fig:ICEYE_locations_10}) constellations. In both figures, optimized ground station locations for unconstrained SCORE, KSAT IP solutions, and Teleport IP solutions are displayed on a global map. The figures display optimized station locations on global maps, with coordinates listed in \Cref{tab:Capella_points} and \Cref{tab:ICEYE}. Translucent circles represent the communication coverage cones for each ground station, calculated for a satellite altitude of 525 km and a minimum elevation angle of $10^{\circ}$.

The maps indicate that orbital inclination strongly influences ground station network design. For the ICEYE constellation, which follows polar or sun-synchronous orbits, optimal ground stations cluster at mid-to-high latitudes to maximize contact opportunities. In contrast, the Capella Space constellation, with mid-inclination orbits, has stations clustered mainly in mid-latitude regions. While latitude primarily determined how often a ground station can see satellites, longitude was found to affect the timing of satellite passes. A wide spread in longitude seemed to help provide more continuous and evenly distributed downlink opportunities over time.  Overall, the combination of latitudinal clustering and longitudinal distribution in the optimized networks demonstrates how both free-placement and fixed-site strategies balance maximizing pass frequency with maintaining temporal coverage.

\input{Capella_points}

\input{ICEYE_points}
\Cref{tab:Capella_points} summarizes metrics for Capella Space ground stations, showing SCORE downlinked 42.43 PB over $T_{opt}$, outperforming KSAT’s 37.44 PB and WTA Teleports’ 41.24 PB. For ICEYE (\Cref{tab:ICEYE}), SCORE achieves 165.20 PB, exceeding KSAT’s 151.06 PB and Teleports’ 148.25 PB by 9–11\%. SCORE consistently selects more diverse, strategically positioned sites in high-latitude or remote areas, leading to substantial data downlink improvements. These results suggest that expanding ground networks beyond traditional infrastructure hubs to underutilized or remote locations can yield meaningful performance gains. However, practical deployment considerations, including backhaul availability and power infrastructure, remain important factors in translating these geometric gains into operational improvements.

\input{appendix_tables}

\subsubsection{Site Feasibility and Infrastructure Proximity Analysis}

To directly address the practical feasibility of free-placement solutions, we analyze the distance between optimized ground station locations and the nearest existing teleport infrastructure in \Cref{fig:teleportFull} for each SCORE variant. \Cref{tab:capella_city_free,tab:iceye_city_free} report coordinates and infrastructure distances for all three variants across both constellations.

Unconstrained SCORE selects geometrically optimal sites averaging 1,011 km (Capella) and 860 km (ICEYE) from the nearest teleport infrastructure, including remote locations such as the South Pole ($-$89.94$^\circ$N) and high-Arctic sites that lack the fiber backhaul and power infrastructure required for commercial EO downlink operations. These results confirm that unconstrained free placement represents a \textit{strong empirical performance benchmark} that can quantify the substantial gains achievable through unconstrained network placement.

The latitude- and infrastructure-constrained variants reveal how much of this gain is recoverable under realistic deployment assumptions. Infrastructure-constrained SCORE reduces mean distance to nearest teleport to 113 km (Capella) and 116 km (ICEYE) --- an order-of-magnitude reduction --- while retaining 96.5\% (41.90/43.43 PB) and 92.0\% (152.03/165.20 PB) of unconstrained performance respectively. The latitude-constrained variant retains 98.0\% (42.58/43.43 PB) for Capella and 93.5\% (154.51/165.20 PB) for ICEYE, at a somewhat higher infrastructure cost (785 km and 810 km mean distance). This demonstrates that substantial geometric gains remain 
achievable through strategic expansion \textit{near} existing infrastructure hubs, without requiring deployment at remote or logistically infeasible sites.

Thus, free placement defines a strong performance benchmark and identifies strategic regions where targeted infrastructure investment could yield significant gains. The constrained variants provide actionable guidance for network planners operating within real-world deployment constraints, bridging the gap between geometric optimality and operational feasibility.

\subsubsection{Additional Antennas versus Ground Stations}

\begin{figure}[htbp!]
    \centering
    \includegraphics[width=\textwidth]{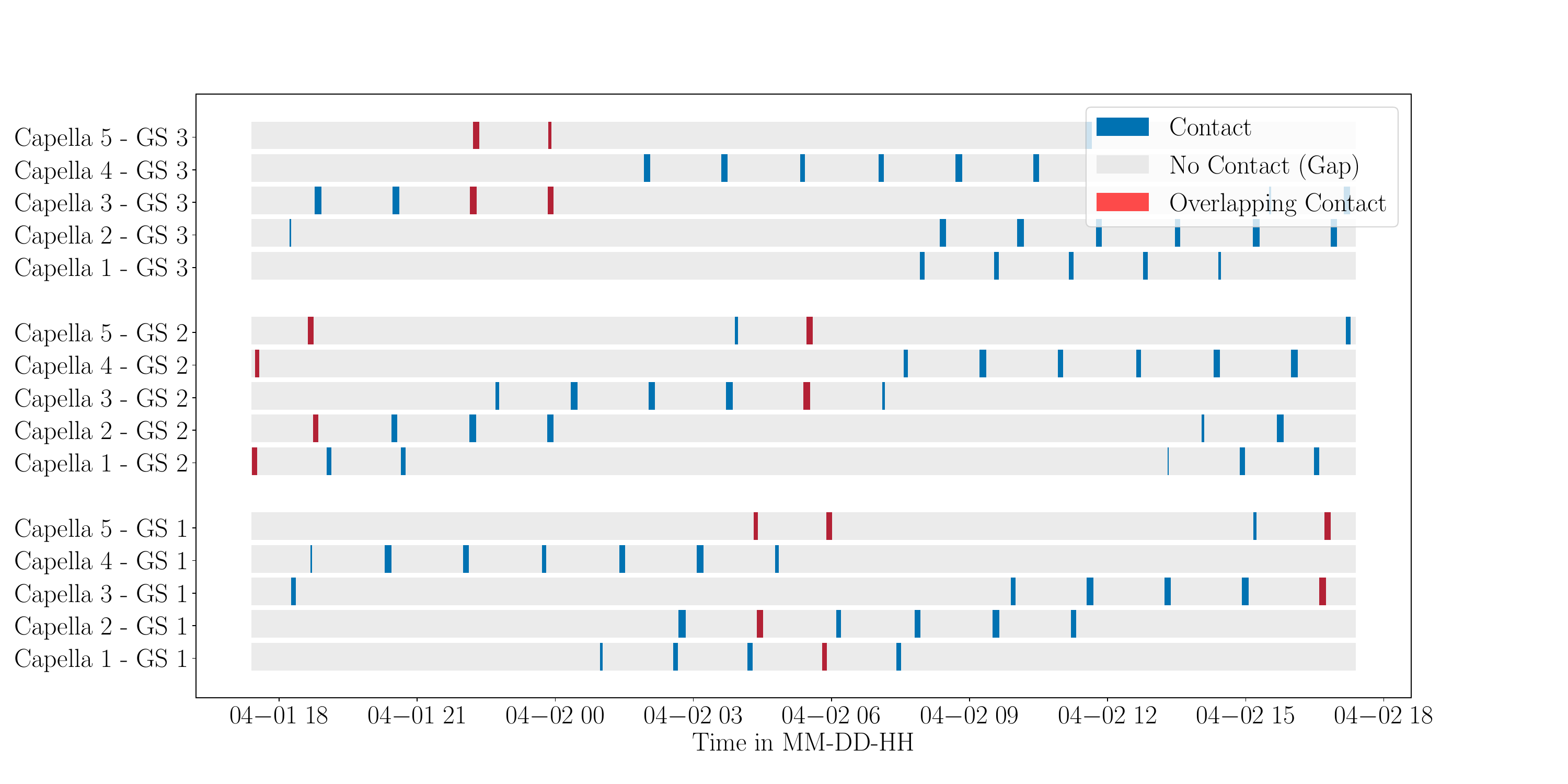} 
   \caption{Contact windows for three optimized SCORE ground stations on Capella: blue (contact), gray (no contact), red (overlapping contacts from multiple satellites).}
  \label{fig:Capella_contactwindows}
\end{figure}

\begin{figure}[htbp!]
    \centering
    \includegraphics[width=\textwidth]{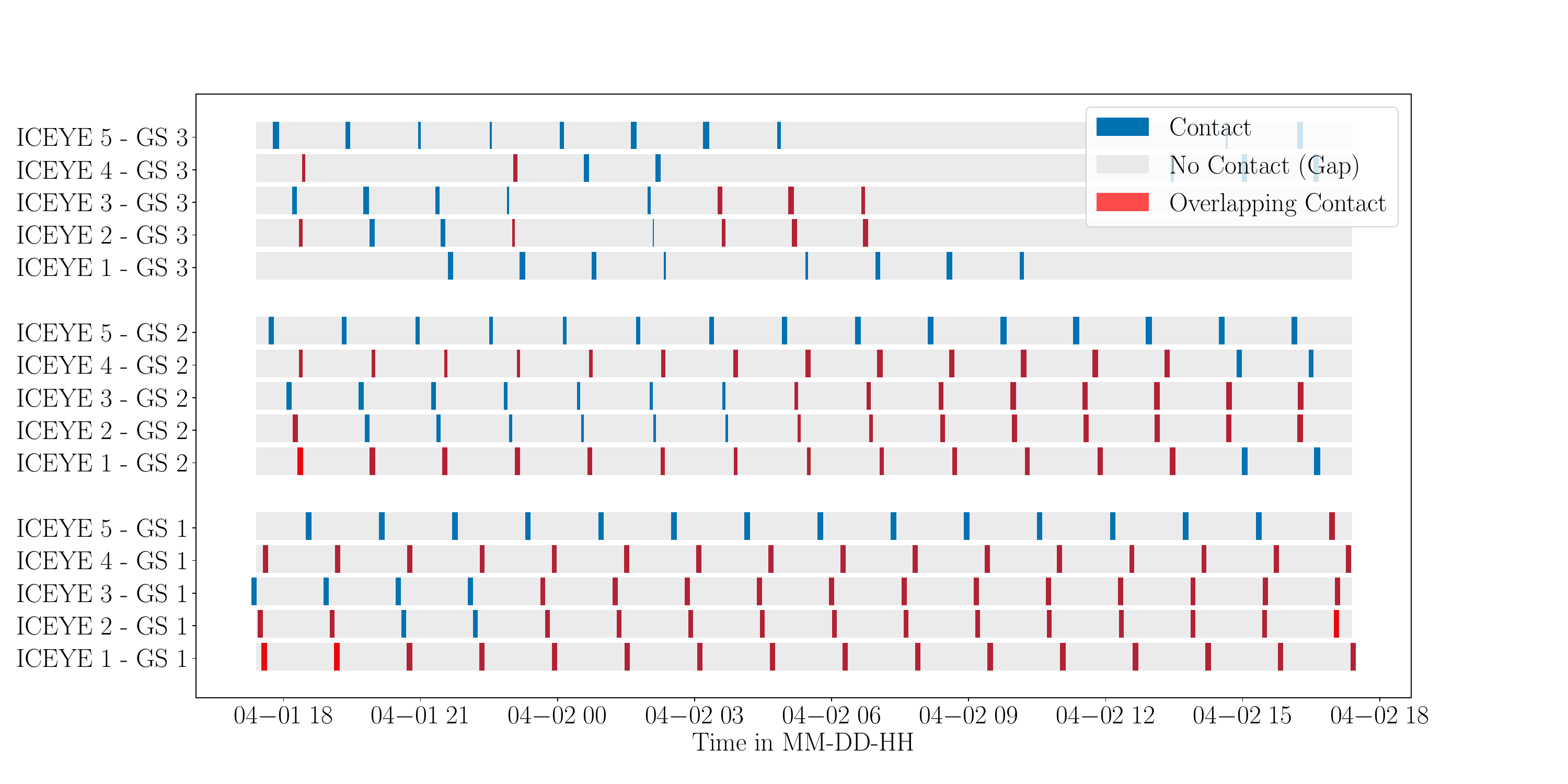} 
   \caption{Contact windows for three optimized SCORE ground stations on the first five ICEYE satellites: blue (contact), gray (no contact), red (overlapping contacts from multiple satellites).}
  \label{fig:ICEYE_contactwindows}
\end{figure}

In all evaluations presented in \Cref{sec:freevsFixed}, each ground station was constrained to one active satellite link at a time, representing a single-antenna configuration. We further investigated the throughput improvements enabled by equipping stations with multiple antennas. \Cref{fig:Capella_contactwindows,fig:ICEYE_contactwindows} display 24-hour contact windows for three example ground stations for the Capella Space and ICEYE constellations, highlighting both utilization patterns and the fundamental limits of single-antenna architectures.

Both constellations exhibit overlapping contact periods, indicated by red regions. For Capella Space (\Cref{fig:Capella_contactwindows}), overlaps recur cyclically across stations. In the ICEYE constellation (\Cref{fig:ICEYE_contactwindows}), overlaps are even more pronounced, especially at high-latitude stations. These overlaps reveal a key limitation of single-antenna stations: simultaneous satellite contacts force prioritization, causing lost downlink opportunities. This challenge intensifies with constellation size, as reflected by substantial red areas in the figures.

The analysis suggests that expanding geographic coverage alone may not significantly increase capacity. Instead, equipping existing stations with multiple antennas at strategic locations offers greater throughput gains. \Cref{tab:contactOpp} quantifies this effect: antenna constraints reduce ICEYE throughput by 68\% at network size 1 (from 85.29 PB to 27.55 PB) and 53\% at network size 20 (from 553.81 PB to 260.89 PB). Capella Space is less impacted, with 7\% and 12\% penalties at network sizes 1 and 20 respectively, reflecting orbital and contact pattern differences. These results demonstrate that for polar-orbit constellations like ICEYE, multi-antenna stations can more than double throughput without increasing geographic footprint, a critical design consideration for satellite network infrastructure.

\input{contactOpp}

%% file: multiSolutions.tex
\begin{table}[htbp!]
\centering
\caption{Aligned ground station coordinates from three optimization cases, reordered to highlight spatial similarity. Overall distribution patterns, not exact sites, primarily drive data downlink performance.}
\label{tab:multi_solution_coordinates}
\begin{tabular}{lrrrrrrrrr}
\toprule
\multicolumn{1}{l}{\phantom{-}} & \multicolumn{2}{c}{Coordinate 1} & \multicolumn{2}{c}{Coordinate 2} & \multicolumn{2}{c}{Coordinate 3} & \multicolumn{2}{c}{Coordinate 4} & 
\multicolumn{1}{l}{\phantom{-}} \\
\cmidrule(lr){2-3} \cmidrule(lr){4-5} \cmidrule(lr){6-7}\cmidrule(lr){8-9}
Scenario & Lon. & Lat. & Lon. & Lat. & Lon. & Lat. & Lon. & Lat. & \textbf{Data (PB)}\\
\midrule
A & $15.65$ & $78.23$ & $2.53$ & $-72.01$ & $-133.72$ & $68.36$ & $-57.85$ & $-51.68$ & $3.11$ \\
B & $-26.51$ & $64.14$ & $2.53$ & $-72.01$ & $-148.49$ & $70.26$ & $168.38$ & $-46.53$ & $3.10$ \\
C & $25.75$ & $71.17$ & $2.53$ & $-72.01$ & $-51.72$ & $64.18$ & $-70.87$ & $-52.94$ & $3.12$ \\
\bottomrule
\end{tabular}
\end{table}

%% file: Capella_points.tex
\begin{table}[bp!]
\centering
\caption{Comparison of KSAT, Teleport, and SCORE ground stations for Capella: 10 stations placed, with SCORE achieving highest data downlink in $T_{\text{opt}}$.}
\label{tab:Capella_points}
\resizebox{\textwidth}{!}{%
\begin{tabular}{@{}rrlrrlrrl@{}}
\toprule
\multicolumn{3}{c}{\textbf{KSAT}} & \multicolumn{3}{c}{\textbf{Teleports}} & \multicolumn{3}{c}{\textbf{SCORE}}\\
\cmidrule(){1-3} \cmidrule(lr){4-6} \cmidrule(){7-9}
\textbf{Lon.} & \textbf{Lat.} & \textbf{Location} & \textbf{Lon.} & \textbf{Lat.} & \textbf{Location} & \textbf{Lon.} & \textbf{Lat.} & \textbf{Location}\\
\cmidrule(){1-3} \cmidrule(lr){4-6} \cmidrule(){7-9}

$-$118.15 & 33.82 & Long Beach, USA & $-$111.95 & 40.78 & Salt Lake City, USA & $-$105.98 & 41.52 & Cheyenne, USA \\
$-$84.26 & 32.95 & Thomaston, USA & $-$52.78 & 47.56 & St. John's, Canada & $-$59.88 & 43.93 & Nova Scotia, Canada \\
$-$25.13 & 36.99 & Azores, Portugal & $-$3.79 & 40.40 & Madrid, Spain & 15.10 & 41.64 & Foggia, Italy \\
22.69 & 38.82 & Thermopylae, Greece & 44.90 & 41.70 & Tbilisi, Georgia & 60.97 & 40.84 & Arap, Turkmenistan\\
$-$70.87 & $-$52.94 & Punta Arenas, Chile & $-$58.31 & $-$34.69 & Buenos Aires, Argentina & $-$62.77 & $-$40.30 & Bahía Blanca, Argentina \\
 27.71 & $-$25.89 & Hartebeesthoek, S. Africa & 18.72 & $-$34.03 & Cape Town, South Africa & 51.73 & $-$46.28 & Alfred Faure, TAAF \\
143.45 & 42.60 & Hokkaido, Japan & 141.26 & 43.17 & Hokkaido, Japan & 107.87 & 41.04 & Bayannur, China \\
$-$156.45 & 20.82 & Maui, USA & $-$157.87 & 21.31 & Honolulu, USA & $-$177.38 & 28.21 & Sand Island, USA \\
168.38 & $-$46.53 & Awarua, New Zealand & 168.38 & $-$46.53 & Awarua, New Zealand & 148.34 & $-$40.39 & Tasmania, Australia \\
115.34 & $-$29.01 & Mingenew, Australia & 115.94 & $-$31.88 & Perth, Australia & 1.92 & $-$89.94 & South Pole \\
\cmidrule(){1-3} \cmidrule(lr){4-6} \cmidrule(){7-9}

\multicolumn{3}{c}{\textbf{37.44 PB Downlinked}} & \multicolumn{3}{c}{\textbf{41.24 PB Downlinked}} & \multicolumn{3}{c}{\textbf{43.43 PB Downlinked}}\\
\bottomrule
\end{tabular}
}
\end{table}

%% file: ICEYE_points.tex
\begin{table}[bp!]
\centering
\caption{Comparison of KSAT, Teleport, and SCORE ground stations for ICEYE: 10 stations placed, with SCORE achieving the highest data downlink in $T_{\text{opt}}$.}
\label{tab:ICEYE}

\resizebox{\textwidth}{!}{%
\begin{tabular}{@{}rrlrrlrrl@{}}
\toprule
\multicolumn{3}{c}{\textbf{KSAT}} & 
\multicolumn{3}{c}{\textbf{Teleports}} & 
\multicolumn{3}{c}{\textbf{SCORE}} \\
\cmidrule(){1-3} \cmidrule(lr){4-6} \cmidrule(){7-9}

\textbf{Lon.} & \textbf{Lat.} & \textbf{Location} & 
\textbf{Lon.} & \textbf{Lat.} & \textbf{Location} & 
\textbf{Lon.} & \textbf{Lat.} & \textbf{Location} \\
\cmidrule(){1-3} \cmidrule(lr){4-6} \cmidrule(){7-9}

$-$133.72 & 68.36 & Inuvik, Canada & 
$-$133.55 & 68.32 & Inuvik, Canada & 
$-$127.32 & 57.78 & British Columbia, Canada \\
$-$148.49 & 70.26 & Prudhoe Bay, USA & 
$-$148.94 & 70.32 & Prudhoe Bay, USA & 
175.48 & 67.33 & Chukotka, Russia \\
15.63 & 78.25 & Svalbard, Norway & 
15.39 & 78.23 & Svalbard, Norway & 
$-$32.32 & 83.55 & Peary Land, Greenland \\
$-$8.62 & 71.00 & Jan Mayen, Norway & 
$-$21.69 & 64.14 & Reykjavík, Iceland & 
$-$6.74 & 58.13 & Scotland, UK \\
31.11 & 70.37 & Vardo, Norway & 
93.51 & 56.25 & Zheleznogorsk, Russia & 
46.22 & 57.47 & Kazan, Russia \\
143.45 & 42.60 & Hokkaido, Japan & 
30.50 & 50.47 & Kiev, Russia & 
102.64 & 66.60 & Chirinda, Russia \\
$-$51.72 & 64.18 & Nuuk, Greenland & 
$-$80.33 & 41.43 & Greenville PA & 
$-$74.13 & 59.81 & Tuttusivik, Canada \\
2.53 & $-$72.01 & Troll, Antarctica & 
2.53 & $-$72.01 & Troll, Antarctica & 
66.52 & $-$90.00 & South Pole \\
$-$67.52 & $-$55.05 & Tolhuin, Argentina & 
$-$67.12 & $-$54.51 & Tolhuin, Argentina & 
$-$67.63 & $-$56.21 & Tolhuin, Argentina \\
168.38 & $-$46.53 & Awarua, New Zealand & 
168.38 & $-$46.53 & Awarua, New Zealand & 
158.96 & $-$55.14 & Macquarie Island, Australia\\
\cmidrule(){1-3} \cmidrule(lr){4-6} \cmidrule(){7-9}
\multicolumn{3}{c}{\textbf{151.06 PB Downlinked}} & \multicolumn{3}{c}{\textbf{148.25 PB Downlinked}} & \multicolumn{3}{c}{\textbf{165.20 PB Downlinked}}\\
\bottomrule
\end{tabular}%
}
\end{table}

%% file: appendix_tables.tex
\begin{table}[h]
\centering
\caption{Comparison of SCORE, SCORE (Lat-Const), and SCORE (Infra-Const) ground station coordinates for Capella Space: 10 stations placed. Distance to the nearest teleport sites in \Cref{fig:teleportFull} is reported for each station.}
\resizebox{0.75\textwidth}{!}{\begin{tabular}{rrcrrcrrc}
\toprule
\multicolumn{3}{c}{\textbf{SCORE}} & \multicolumn{3}{c}{\textbf{SCORE (Lat-Const)}} & \multicolumn{3}{c}{\textbf{SCORE (Infra-Const)}} \\
\cmidrule(lr){1-3} \cmidrule(lr){4-6} \cmidrule(lr){7-9}
\textbf{Lon.} & \textbf{Lat.} & \textbf{Dist. (km)} & \textbf{Lon.} & \textbf{Lat.} & \textbf{Dist. (km)}  & \textbf{Lon.} & \textbf{Lat.} & \textbf{Dist. (km)} \\
\cmidrule(lr){1-3} \cmidrule(lr){4-6} \cmidrule(lr){7-9}
107.87     & 41.04  & 713  & $-$8.14    & 41.03  & 259  & $-$159.61 & 21.34  & 143 \\
$-$105.98  & 41.52  & 112  & 68.77      & $-$48.50 & 4155 & $-$166.29 & 53.35  & 61  \\
148.34     & $-$40.39 & 292 & $-$176.95 & 51.49  & 750  & 140.89    & 41.99  & 106 \\
$-$177.38  & 28.21  & 2063 & 173.02     & $-$40.77 & 176 & $-$58.88  & $-$33.72 & 78 \\
15.10      & 41.64  & 44   & 109.72     & 41.01  & 558  & 136.68    & $-$36.17 & 231 \\
51.73      & $-$46.28 & 3055 & $-$109.69 & 40.50 & 118  & $-$124.39 & 40.25  & 239 \\
1.92       & $-$89.94 & 1993 & 136.68    & $-$36.17 & 231 & 19.53    & $-$34.77 & 102 \\
$-$62.77   & $-$40.30 & 487 & $-$62.79  & $-$40.32 & 490 & $-$81.38 & 40.98  & 11  \\
$-$59.88   & 43.93  & 383  & 20.00      & $-$34.91 & 144 & $-$4.41  & 51.00  & 119 \\
60.97      & 40.84  & 968  & 60.97      & 40.84  & 968  & 44.52    & 41.45  & 42  \\
\cmidrule(){1-3}\cmidrule(lr){4-6}\cmidrule(lr){7-9}
\multicolumn{2}{r}{\textbf{Mean dist. (km)}} & \multicolumn{1}{c}{\textbf{1011}} & 
\multicolumn{2}{r}{\textbf{Mean dist. (km)}} & \multicolumn{1}{c}{\textbf{785}} & 
\multicolumn{2}{r}{\textbf{Mean dist. (km)}} & \multicolumn{1}{c}{\textbf{113}} \\
\multicolumn{2}{r}{\textbf{Max dist. (km)}} & \multicolumn{1}{c}{\textbf{3055}} & 
\multicolumn{2}{r}{\textbf{Max dist. (km)}} & \multicolumn{1}{c}{\textbf{4155}} & 
\multicolumn{2}{r}{\textbf{Max dist. (km)}} & \multicolumn{1}{c}{\textbf{239}} \\
\cmidrule(){1-3}\cmidrule(lr){4-6}\cmidrule(lr){7-9}
\multicolumn{3}{c}{\textbf{43.43 PB Downlinked}} & \multicolumn{3}{c}{\textbf{42.58 PB Downlinked}} &
\multicolumn{3}{c}{\textbf{41.90 PB Downlinked}}\\
\bottomrule
\end{tabular}}
\label{tab:capella_city_free}
\end{table}

\begin{table}[h]
\centering
\caption{Comparison of SCORE, SCORE (Lat-Const), and SCORE (Infra-Const) ground station coordinates for ICEYE: 10 stations placed. Distance to the nearest teleport sites in \Cref{fig:teleportFull} is reported for each station.}
\resizebox{0.75\textwidth}{!}{\begin{tabular}{rrcrrcrrc}
\toprule
\multicolumn{3}{c}{\textbf{SCORE}} & \multicolumn{3}{c}{\textbf{SCORE (Lat-Const)}} & \multicolumn{3}{c}{\textbf{SCORE (Infra-Const)}} \\
\cmidrule(lr){1-3} \cmidrule(lr){4-6} \cmidrule(lr){7-9}
\textbf{Lon.} & \textbf{Lat.} & \textbf{Dist. (km)} & \textbf{Lon.} & \textbf{Lat.} & \textbf{Dist. (km)}  & \textbf{Lon.} & \textbf{Lat.} & \textbf{Dist. (km)} \\
\cmidrule(lr){1-3} \cmidrule(lr){4-6} \cmidrule(lr){7-9}
$-$127.32  & 57.78   & 379   & 2.73       & $-$72.5 & 58    & 2.53       & $-$72.01 & 0   \\
66.52      & $-$90.00 & 2000 & 11.69      & 79.07   & 123   & 166.54     & $-$48.27 & 238 \\
$-$32.32   & 85.35   & 1033  & $-$113.62  & 70.74   & 816   & 12.69      & 79.07   & 110  \\
$-$74.13   & 59.81   & 1110  & 166.56     & $-$48.26 & 236  & 148.49     &   70.26 & 15  \\
158.96     & $-$55.14 & 1161 & 150.00     & 72.27   & 1794  & $-$3.82    & 52.06   & 67   \\
102.64     & 66.60   & 953   & 82.46      & 61.40   & 637   & 26.11    & 70.20 & 270 \\
$-$8.13    & 58.19   & 357   & $-$67.20   & $-$56.25 & 193  & $-$133.58  & 69.02   & 78   \\
175.48     & 67.33   & 917   & $-$7.35    & 52.96   & 90    & 139.11     & 36.61   & 102  \\
46.22      & 57.47   & 500   & 70.11      & $-$49.97 & 4057 & $-$61.63   & 53.08   & 90   \\
$-$67.63   & $-$56.21 & 192  & $-$164.28  & 53.45   & 156   & $-$67.20   & $-$56.25 & 193 \\
\cmidrule(){1-3}\cmidrule(lr){4-6}\cmidrule(lr){7-9}
\multicolumn{2}{r}{\textbf{Mean dist. (km)}} & \multicolumn{1}{c}{\textbf{860}} & 
\multicolumn{2}{r}{\textbf{Mean dist. (km)}} & \multicolumn{1}{c}{\textbf{810}} & 
\multicolumn{2}{r}{\textbf{Mean dist. (km)}} & \multicolumn{1}{c}{\textbf{116}} \\
\multicolumn{2}{r}{\textbf{Max dist. (km)}} & \multicolumn{1}{c}{\textbf{2000}} & 
\multicolumn{2}{r}{\textbf{Max dist. (km)}} & \multicolumn{1}{c}{\textbf{4057}} & 
\multicolumn{2}{r}{\textbf{Max dist. (km)}} & \multicolumn{1}{c}{\textbf{270}} \\
\cmidrule(){1-3}\cmidrule(lr){4-6}\cmidrule(lr){7-9}
\multicolumn{3}{c}{\textbf{165.20 PB Downlinked}} & \multicolumn{3}{c}{\textbf{154.51 PB Downlinked}} &
\multicolumn{3}{c}{\textbf{152.03 PB Downlinked}}\\
\bottomrule
\end{tabular}}
\label{tab:iceye_city_free}
\end{table}

%% file: contactOpp.tex
\begin{table}[htbp!]
\centering
\caption{Comparison of total data downlinked between single-antenna and multi-antenna SCORE for supporting Capella Space and ICEYE constellations.}
\label{tab:contactOpp}

\resizebox{0.9\textwidth}{!}{%
\begin{tabular}{rrrrr}
\toprule

\multicolumn{1}{l}{\phantom{-}} & \multicolumn{4}{c}{\textbf{SCORE}}\\
\cmidrule(lr){2-5}
\multicolumn{1}{l}{\phantom{-}} & \multicolumn{2}{c}{\textbf{Capella Space}} & \multicolumn{2}{c}{\textbf{ICEYE}} \\
\cmidrule(lr){2-3} \cmidrule(lr){4-5}
Network Size & \textbf{No Antenna Constraint} & Antenna Constraint & \textbf{No Antenna Constraint} & Antenna Constraint\\
\midrule
(1) & 5.04 PB & 4.70 PB & 85.29 PB & 27.55 PB \\
(2) & 10.08 PB & 9.23 PB & 161.76 PB & 53.75 PB \\
(3) & 15.08 PB & 13.68 PB & 200.63 PB & 70.70 PB \\
(4) & 19.94 PB & 18.04 PB & 239.48 PB & 87.75 PB \\
(5) & 25.10 PB & 22.76 PB & 259.67 PB & 99.65 PB \\
(7) & 34.08 PB & 30.78 PB & 310.95 PB & 125.99 PB \\
(10) & 46.35 PB & 42.41 PB & 382.38 PB & 165.20 PB \\
(15) & 66.37 PB & 58.55 PB & 470.65 PB & 214.42 PB \\
(20) & 82.15 PB & 72.44 PB & 553.81 PB & 260.89 PB \\
\bottomrule
\end{tabular}
}
\end{table}

%% file: 010_Conclusion.tex
\section{Conclusion}
In this work, we introduced SCORE, an algorithm for ground station network design that supports free-placement location optimization. SCORE employs a two-step process, sequential coordinate selection followed by sequential cyclic refinement, to effectively address high-dimensional optimization challenges, enabling the design of larger ground station networks to support increasingly complex satellite constellations.

We conducted a comprehensive evaluation of SCORE against both single-step free-placement optimization methods, such as differential evolution, and fixed-site selection optimization IP baselines using real-world infrastructure from KSAT and the World Teleport Association. In the free-placement setting, SCORE consistently outperformed DE in both computational efficiency and solution quality. Across our studies, SCORE required $5\times$ fewer function evaluations than DE to reach convergence, while achieving 13\% higher data downlink throughput. These results highlight that SCORE’s sequential coordinate selection and local refinement strategy effectively overcomes DE’s limitations in fine-grained exploitation, as our method not only converges faster but also finds higher-quality station placements. Our results also underscore that the overall composition of station locations plays a more critical role than any individual site in achieving robust satellite communication and data downlink performance.

When compared to fixed-site selection optimizations, where IP-based formulations yield optimal solutions within a limited set of locations, SCORE's unconstrained global search establishes a strong empirical performance benchmark, achieving up to 15\% greater total data downlink under flexible siting conditions. This quantifies the geometric performance potential beyond pre-existing site constraints and motivates strategic infrastructure expansion in high-value regions. Infrastructure-constrained SCORE retains over 92\% of this gain while remaining near existing fiber and power infrastructure, demonstrating that substantial improvements are achievable within realistic deployment constraints. We also examined the trade-offs between adding antennas to existing stations and expanding the network with strategically placed new ground stations.

SCORE offers a practical solution for satellite operators aiming to optimize their ground infrastructure, while also opening new directions for research in scalable and adaptive network design. While developed for free-placement ground station optimization, SCORE's approach is broadly applicable to other problems requiring structured search over complex configuration spaces. As satellite constellations expand in both size and complexity, methods like SCORE will be essential for efficiently evaluating and designing ground station networks that can effectively support increasing data demands and evolving mission requirements of growing satellite constellations. The complete codebase supporting this study is publicly available at: \hyperlink{https://github.com/sisl/loc-gsopt}{https://github.com/sisl/loc-gsopt}.

%% file: 011_Appendix.tex
\section*{Appendix}



\subsection{Alternative Objective Function: Mean Gap-Time Minimization}
\label{app:meangap}
Another possible objective includes \textit{mean gap time minimization}, which seeks to reduce the average time between consecutive contacts in order to prevent prolonged communication gaps for any satellite. For an EO mission, this encourages a more consistent communication cadence across the constellation by reducing the typical downtime between uplink and downlink opportunities. For the purpose of computing mean gap times, only feasible contact opportunities are considered (i.e., for each contact pair $(c,C)$, we set $c=1$ to indicate that every feasible contact is included in the gap calculation). This ensures the gap metric measures the inherent communication cadence of the network, independent of scheduling decisions.
Let $\mathcal{C}_S\subseteq \mathcal{C}_{\mathcal{L},\mathcal{S}}$ be the contacts for a satellite $S\in\mathcal{S}$ supported by ground network $\mathcal{L}$. We assume all contacts $ (c,C) \in \mathcal{C}_S $ are ordered in ascending order by the start time $ C^{\text{start}} $. This ordering allows us to define communication gaps as the difference $ C_{i+1}^{\text{start}} - C_i^{\text{end}} $, where $ C_i^{\text{end}} \leq C_{i+1}^{\text{start}} $. If $ C_i^{\text{end}} \geq C_{i+1}^{\text{start}} $, the previous contact window overlaps with or ends exactly at the start of the next contact window, resulting in continuous communication with no gaps between contacts.
 For the length of all contacts $i =1,\dots,\lvert\mathcal{C}_S\rvert-1$, we define the positive gaps lengths $g_{\text{gaps}}(\mathcal{L},S)$ for each satellite as:
\begin{equation}
\begin{aligned}
    g_{\text{gaps}}(\mathcal{L},S) = \max(0,C_{i+1}^{\text{start}} - C_{i}^{\text{end}})
\end{aligned}
\label{eqn:gap_list}
\end{equation}
from contacts generated with a given candidate network $\mathcal{L}$. The number of real gaps for satellite $S$ is then defined as 
\begin{equation}
\begin{aligned}
    n_{\text{gaps}}(\mathcal{L},S) = \sum_{S\in\mathcal{S}}\textbf{1}\{g_{\text{gaps}}(\mathcal{L},S)>0\}
\end{aligned}
\label{eqn:gap_length}
\end{equation} where only gaps larger than zero seconds are counted. Using both \Cref{eqn:gap_list} and \Cref{eqn:gap_length}, the per-satellite mean gap time is then calculated as 
\begin{equation}
        \bar{g}_{\text{gaps}}(\mathcal{L},S) = 
    \begin{cases}
    \begin{aligned}
        \dfrac{\sum_{S \in\mathcal{S}} g_{\text{gaps}}(\mathcal{L},S)}{n_{\text{gaps}}(\mathcal{L},S)},  &&n_{\text{gaps}}(\mathcal{L},S) >0 \\
    0\quad\quad\quad, && n_{\text{gaps}}(\mathcal{L},S) =0
    
    \end{aligned}
    \end{cases}
\end{equation}
for both when $n$ is non-zero, or when there exist no gaps within the mission window. This ensures the mean is computed over actual communication gaps, excluding overlapping contacts. The full objective function then becomes
\begin{equation}
\begin{aligned}
    f_{\text{gap}}(\mathcal{L,S}) = \frac{T_{\text{opt}}}{T_{\text{sim}}}\frac{1}{\lvert\mathcal{S}\rvert}
    \sum_{S \in {\mathcal{S}}}
    \bar{g}_{\text{gaps}}(\mathcal{L},S)
\end{aligned}
\label{eqn:obj_min_gap}
\end{equation}
which represents the mean length of communication gap times for each satellite, averaged over the entire constellation. Once again, we weight the objective by $\frac{T_{\text{opt}}}{T_{\text{sim}}}$ to appropriately reflect the expected total downtime over the entire mission.

\subsection{Synthetic Walker-Star Constellations}
We outline how to calculate several parameters necessary to define a synthetic Walker-Star constellation for simulation purposes, including the right ascension of the ascending node for each orbital plane, the mean anomaly of each satellite to introduce phasing, and a unique indexing scheme to identify satellites within and across planes. The Right Ascension of the Ascending Node ($\Omega$) for each plane $\rho \in \{1,\dots,N_{\rho}\}$ is uniformly distributed across $360\degree$ as
\begin{equation}
    \Omega_\rho = \rho \cdot \frac{360}{N_{\rho}}.
    \label{eqn:raan}
\end{equation}

Each satellite $S \in \mathcal{S}$ in this Walker-Star constellation is uniquely indexed by its orbital plane $\rho$ and its in-plane index $s$, written as $S_{\rho,s}$. To introduce phasing between planes, we define the mean anomaly $M$ of satellite $S_{\rho,s}$ as
\begin{equation}
    M_{\rho,s} = \frac{360}{s_{\text{per plane}}}\,s + \frac{720\,\rho}{|\mathcal{S}|},
    \label{eqn:meanAnomaly}
\end{equation}
where $s = 1,\dots,s_{\text{per plane}}$ is the in-plane satellite index and $s_{\text{per plane}}$ is the number of satellites per orbital plane. The first term in \Cref{eqn:meanAnomaly} governs the relative placement of satellites within each plane, while the second term introduces inter-plane phasing. For this synthetic Walker-Star constellation, since $s_{\text{per plane}} = 1$, the in-plane spacing term vanishes, and $M_{\rho,s}$ is determined entirely by inter-plane phasing.

\subsection{Surrogate Optimization Analysis}
\label{app:SurrgOpt}

As mentioned in \cref{sec:ProblemFormulation}, long-term orbital propagation cannot reliably predict contact opportunities with sufficient accuracy over multi-year timespans, as stochastic perturbations experienced by LEO spacecraft, specifically solar radiation pressure and atmospheric drag, accumulate over time. To capture the cyclic nature of satellite contacts across orbital cycles, we follow prior work~\cite{eddy_optimal_2024,kim2026scalable} and adopt a 7--10 day surrogate window as a statistically representative approximation of long-term contact distributions.

To validate this assumption, we performed a parameter sweep of 5,250 configurations across varying altitudes, inclinations, and ground station locations, detailed in \Cref{tab:sweep_params}. For each configuration, contacts were simulated at 365 discrete durations (1-day, 2-day, \ldots, 365-day windows), yielding a running mean of contacts/day and duration/day at each duration relative to the 365-day reference. Of the 5,250 configurations, 3,644 yielded non-zero contacts and are included in the analysis.

\begin{table}[h]
\centering
\caption{Surrogate validation parameter sweep.}
\begin{tabular}{lr}
\toprule
Parameter & Range \\
\midrule
Altitude & 300--1000~km (50~km steps) \\
Inclination & 0--90$^\circ$ (10$^\circ$ steps) \\
KSAT locations & Stations from the 35 ground station set \\
Simulation durations & 1--365 days (1-day steps) \\
\midrule
\textbf{Total configs} & \textbf{5,250 (3,644 non-zero contacts)} \\
\bottomrule
\end{tabular}
\label{tab:sweep_params}
\end{table}

\begin{figure}[htbp]
    \centering
    \includegraphics[width=\linewidth]{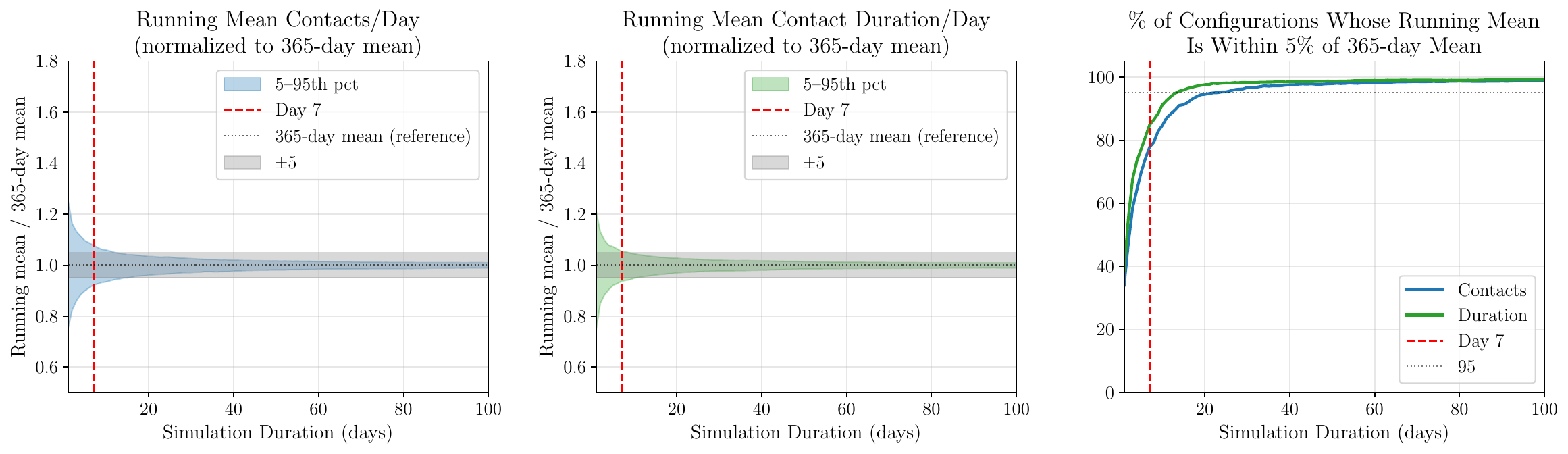}
    \caption{Convergence of running mean contact statistics to the 365-day reference across 3,644 configurations. The red dashed line marks the adopted 7-day surrogate window.}
    \label{fig:surrogate}
\end{figure}

\Cref{fig:surrogate} presents running means of contacts/day and duration/day normalized to 365-day references, shown over the first 100 days for clarity. The running mean contact statistics converge steadily toward the full-year mean as simulation duration increases. A 7-day surrogate window estimates annual mean contact frequency and duration within 5\% for over 70\% of configurations tested, rising to over 95\% with a longer window of 20--25 days. Given that propagation accuracy degrades at longer horizons due to accumulated stochastic perturbations, we adopt 7--10 days as a practical surrogate duration balancing statistical representativeness with propagation fidelity. These results substantially extend the preliminary analysis of~\citeauthor{eddy_optimal_2024}~\cite{eddy_optimal_2024}, providing large-scale empirical justification for this assumption.

\subsection{Differential Evolution Hyperparameter Sweep}
\label{app:DESweep}
\begin{figure}
    \centering
    \includegraphics[width=\linewidth]{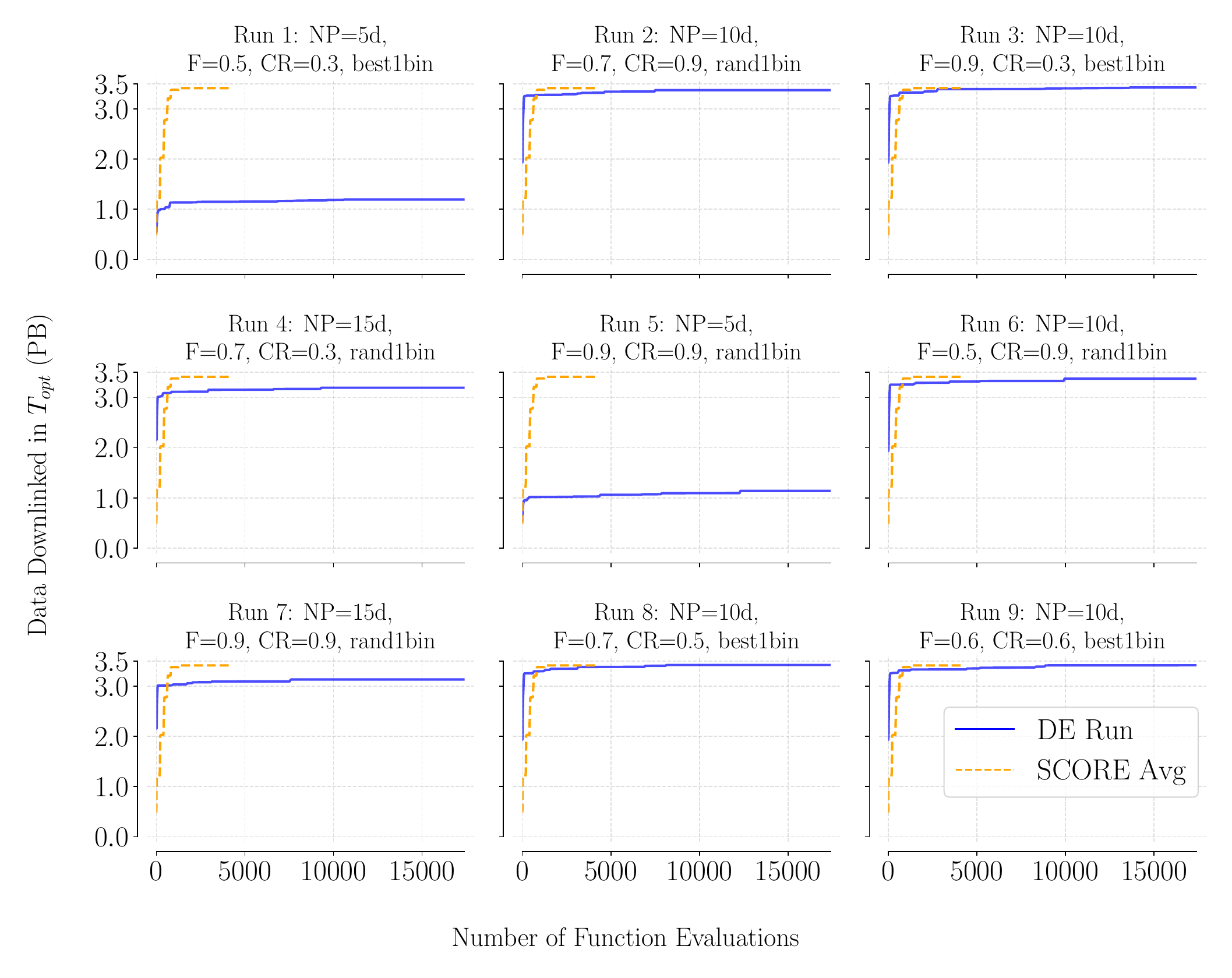}
    \caption{Convergence comparison of SCORE versus 9 DE hyperparameter configurations: SCORE converges $4 $ to $5\times$ faster, with several DE runs failing to converge within the evaluation budget.}
    \label{fig:DEsweeps}
\end{figure}

To ensure a fair comparison between SCORE and Differential Evolution (DE), we conducted a hyperparameter sweep over 9 configurations for our largest synthetic Walker-Star constellation experiment in \Cref{sec:SCOREVSDE}. The sweep spanned crossover rate $CR \in \{0.3, 0.5, 0.9\}$, mutation scale $F \in \{0.5, 0.6, 0.7, 0.9\}$, population size $NP \in \{5, 10, 15\}$, and mutation strategy $\in \{\texttt{best1bin}, \texttt{rand1bin}\}$. We note the final selected configurations in \Cref{tab:de_params}. These ranges were selected to cover both conservative and aggressive exploration strategies, following standard recommendations from the DE literature~\cite{storn_differential_1997}. We note that while the majority of DE configurations can eventually approach SCORE's solution quality given sufficient function evaluations, identifying well-performing hyperparameters itself requires additional tuning effort. The convergence behavior of all configurations is shown in \Cref{fig:DEsweeps}. Despite this broad sweep, no configuration matched SCORE's convergence speed, with most requiring 4–5$\times$ more function evaluations to reach comparable solution quality, and two configurations (Runs 1 and 5) failing to converge to competitive solutions entirely.

\begin{table}[h]
\centering
\caption{Differential Evolution hyperparameter configurations tested in parameter sweep. Population size is scaled by problem dimensionality $d$ (e.g., $5d$ means $5 \times d$ where $d = 2n$ for $n$ ground stations).}
\begin{tabular}{cccccl}
\toprule
Run & Pop. Size & $F$ & CR & Strategy & Purpose \\
\midrule
1  & $5d$  & 0.5 & 0.3 & \texttt{best1bin} & Fast convergence, local exploitation \\
2  & $10d$ & 0.7 & 0.9 & \texttt{rand1bin} & Balanced exploration \\
3  & $10d$ & 0.9 & 0.3 & \texttt{best1bin} & High mutation, low mixing \\
4  & $15d$ & 0.7 & 0.3 & \texttt{rand1bin} & Conservative large population \\
5  & $5d$  & 0.9 & 0.9 & \texttt{rand1bin} & Chaotic, high exploration \\
6  & $10d$ & 0.5 & 0.9 & \texttt{rand1bin} & Baseline (from original experiments)\\
7  & $15d$ & 0.9 & 0.9 & \texttt{rand1bin} & Maximum exploration, large pop \\
8  & $10d$ & 0.7 & 0.5 & \texttt{best1bin} & Moderate exploration \\
9  & $10d$ & 0.6 & 0.6 & \texttt{best1bin} & Moderate balanced configuration \\
\bottomrule
\end{tabular}
\label{tab:de_params}
\end{table}

This hyperparameter sensitivity becomes increasingly problematic as ground station network sizes grow, since each function evaluation requires propagating satellite orbits and computing access windows over the full simulation window, making exhaustive parameter sweeps computationally prohibitive at scale. In contrast, SCORE requires no hyperparameter tuning beyond the choice of local optimizer, and consistently converges to near-optimal solutions with modest growth in function evaluations as network size increases.

\subsection{SCORE Cyclic Refinement Ablation Study}
\label{app:score_ablation}

We provide an ablation study of the cyclic refinement step in the SCORE algorithm, addressing two questions: (1) whether the cyclic refinement step provides meaningful performance gains over greedy-only placement, and (2) whether randomizing the ground station selection order within the refinement step affects solution quality.

\subsubsection{Effect of Cyclic Refinement}

The SCORE algorithm consists of two phases: an initial greedy placement phase, in which ground stations are selected sequentially to maximize marginal coverage, followed by a cyclic refinement phase, in which each station is re-optimized while holding all others fixed, repeated for $E_{\text{max}}$ cycles. To assess the contribution of the refinement phase, we compare convergence behavior with and without cyclic refinement enabled across the four-satellite, four-station Walker-Star scenario described in \Cref{sec:syntheticWalkerStar}.

As shown in \Cref{fig:score_ablation}, the vertical dashed line at approximately 800 function evaluations marks the boundary between the greedy placement phase and the onset of cyclic refinement. The greedy phase alone achieves a substantial fraction of the final solution quality; however, the cyclic refinement phase consistently yields additional improvement, particularly in runs where the greedy placement produced a suboptimal initial configuration. This demonstrates that the second phase is not redundant and contributes measurably to final solution quality.

\subsubsection{Effect of Randomized Selection Order}

We additionally investigate whether randomizing the order in which ground stations are selected for re-optimization during the cyclic refinement phase affects performance. For each of nine independent trials with distinct random seeds, we run SCORE with both a fixed deterministic ordering ($i = 0, 1, \ldots, n-1$) and a randomly shuffled ordering, where the shuffle is re-drawn at each refinement cycle using a seeded random number generator.

\begin{figure}[h]
    \centering
    \includegraphics[width=\textwidth]{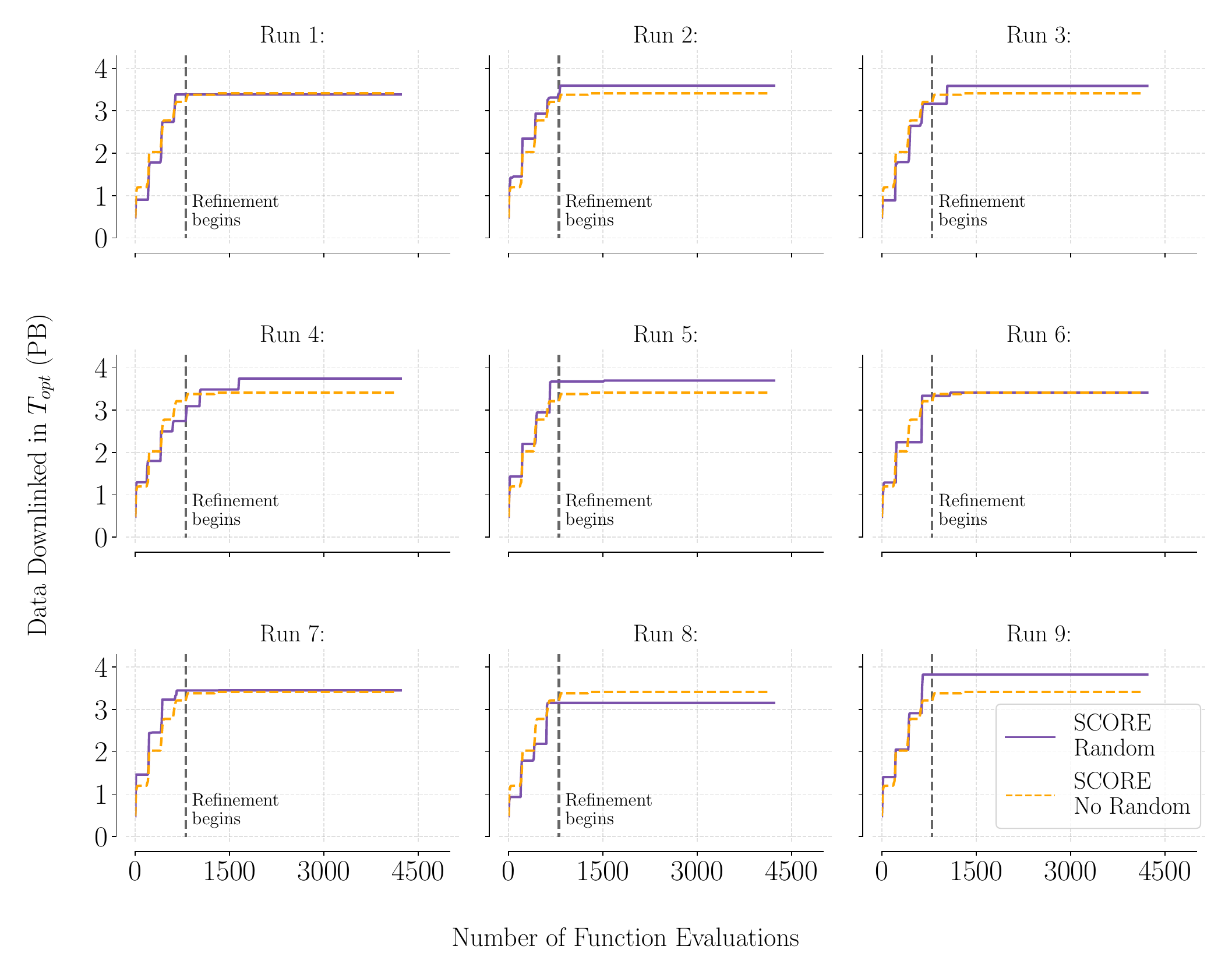}
    \caption{SCORE solution quality is robust to cyclic refinement ordering: randomized and deterministic variants converge comparably across nine independent Walker-Star trials.}
    \label{fig:score_ablation}
\end{figure}

\Cref{fig:score_ablation} shows the convergence curves for both conditions across all nine trials. In the majority of runs, the randomized and deterministic orderings converge to comparable final solutions, with neither consistently outperforming the other. Some runs exhibit transient differences in convergence trajectory, but these differences do not persist to the final solution. This suggests that SCORE's performance is robust to the specific selection order used in the refinement step, and that the deterministic ordering used in the primary results is not a source of systematic bias.

\section*{Funding Sources}

This research was supported by the Hertz Foundation and the National Defense Science and Engineering Graduate (NDSEG) Fellowship Program.

